 \documentclass[12pt,preprint2]{aastex63}

\usepackage{multirow}

\usepackage[utf8]{inputenc}
\usepackage{natbib}
\usepackage{graphicx}
\usepackage{longtable}
\usepackage{enumerate}
\usepackage{amsmath}
\usepackage{hyperref}
\usepackage{CJK}
\usepackage[title]{appendix}
\usepackage{rotating}
\usepackage{longtable}

\def\msun{\ifmmode {\rm M}_{\mathord\odot}\else $M_{\mathord\odot}$\fi}
\def\rsun{\ifmmode {\rm R}_{\mathord\odot}\else $R_{\mathord\odot}$\fi}
\def\lsun{\ifmmode {\rm L}_{\mathord\odot}\else $L_{\mathord\odot}$\fi}

\def\co{$^{12}$CO}
\def\c18o{C$^{18}$O}
\def\h2{H$_{2}$}
\def\13co{$^{13}$CO}
\def\n2hp{$_{2}$H$^{+}$}

\def\cm2{cm$^{-2}$}

\def\kkms{K$\cdot$km/s}

\newcommand{\kms}{km~s$^{-1}$}

\newcommand{\CASI}{{\sc casi}}
\newcommand{\CASItD}{{\sc casi-3d}}

\shorttitle{}
\shortauthors{}

\begin{document}
\begin{CJK*}{UTF8}{gbsn}

\title{A Census of Protostellar Outflows in Nearby Molecular Clouds}

\author{Duo Xu}
\affil{Department of Astronomy, The University of Texas at Austin, Austin, TX 78712, USA;}
\affil{Department of Astronomy, University of Virginia, Charlottesville, VA 22904-4235, USA;}
\email{xuduo117@utexas.edu}

\author{Stella S. R. Offner}
\affil{Department of Astronomy, The University of Texas at Austin, Austin, TX 78712, USA;}
\email{soffner@astro.as.utexas.edu}

\author{Robert Gutermuth}
\affil{Department of Astronomy, University of Massachusetts, Amherst, MA 01003, USA;}

\author{Shuo Kong}
\affil{Steward Observatory, University of Arizona, Tucson, AZ 85719, USA;}
\affil{Department of Astronomy, Yale University, New Haven, CT 06511, USA;}

\author{Hector G. Arce}
\affil{Department of Astronomy, Yale University, New Haven, CT 06511, USA;}


\begin{abstract}
We adopt the deep learning method \CASItD\ (Convolutional Approach to Structure Identification-3D) to systemically identify protostellar outflows in $^{12}$CO and $^{13}$CO observations of the nearby molecular clouds, Ophiuchus, Taurus, Perseus and Orion. The total outflow masses are 267 \msun, 795 \msun, 1305 \msun\ and 6332 \msun\ for Ophiuchus, Taurus, Perseus and Orion, respectively. We show the outflow mass in each cloud is linearly proportional to the total number of young stellar objects. The estimated total 3D { deprojected} outflow energies are $9\times10^{45}$ ergs, $6\times10^{46}$ ergs, $1.2\times10^{47}$ ergs and $6\times10^{47}$ ergs for Ophiuchus, Taurus, Perseus and Orion, respectively. The energy associated with outflows is sufficient to offset turbulent dissipation at the current epoch for all four clouds. All clouds also exhibit a break point in the spatial power spectrum of the outflow prediction map, which likely corresponds to the typical outflow mass and energy injection scale. 


\end{abstract}


\section{Introduction} 

Protostars launch energetic collimated bipolar outflows during the star formation process. This outflow gas entrains and accelerates the surrounding material to higher velocities than that of the ambient cloud gas, injecting a substantial amount of energy into the host molecular cloud \citep{2014prpl.conf..451F, 2016ARA&A..54..491B}. 
Numerical simulations suggest protostellar outflows not only reduce the core and stellar mass \citep[e.g.,][]{2012ApJ...747...22H,2017ApJ...847..104O} but also depress the star formation rate and star formation efficiency in molecular clouds \citep[e.g.,][]{2014ApJ...790..128F,2015MNRAS.450.4035F,2018MNRAS.476..771C}. Meanwhile, protostellar outflows likely control the shape of the stellar initial mass function \citep{2018MNRAS.476..771C,2021MNRAS.502.3646G}. Theoretical and numerical studies indicate that protostellar outflows are highly efficient at
driving turbulent motions \citep{2007ApJ...659.1394M,2007ApJ...662..395N,2013MNRAS.432L..80M}. In small to intermediate size clouds, energy injected by protostellar outflows can compensate for rapid turbulent dissipation over at least several dynamical time scales \citep{2006ApJ...653..416C,2006ApJ...640L.187L,2007ApJ...659.1394M,2009ApJ...695.1376C}. Consequently, protostellar outflows may stall the local gravitational collapse of molecular clouds, thereby extending the cloud lifetime \citep{2010ApJ...709...27W}. 
The role of protostellar feedback on the fate of molecular clouds is still under debate \citep{2016ARA&A..54..491B}. 

Unfortunately, the mechanism by which protostellar outflows convert their kinetic energy into turbulent energy remains poorly understood observationally \citep{2014prpl.conf..451F}. In practice, astronomers usually evaluate outflow impact by comparing the total kinetic energy from protostellar outflows with the turbulent energy of their host cloud. For example, \citet{2010ApJ...715.1170A} identified 60 outflow candidates in the Perseus molecular cloud and concluded the total outflow energy is sufficient to replenish the dissipation of turbulence. A similar result was also found in Taurus \citep{2015ApJS..219...20L}, in Ophiuchus \citep{2011ApJ...726...46N}, and in Orion \citep{2020ApJ...896...11F}. 

Accurately determining the impact of protostellar outflows,
especially quantifying their effect on the cloud energy budget, requires a complete census of outflows. However, it is challenging to identify most of the outflows that are deeply embedded in dense clouds \citep{2010ApJ...715.1170A, 2014ApJ...783...29D}. The morphology of these embedded outflows are not as distinct as isolated ones in low velcity channels 
and the outflowing gas is difficult to separate from the ambient cloud. Protostars are often clustered and their outflows interact, making visual identification challenging. One solution is to calculate the outflow gas whose velocity is significantly above the cloud velocity dispersion and extrapolate the total mass \citep{2010ApJ...715.1170A,2015ApJS..219...20L,2020ApJ...896...11F}. However, this leads to an order of magnitude uncertainty \citep{2010ApJ...715.1170A}. 

Recent developments in machine learning approaches have enabled automated detection, which can separate outflows from cloud emission and conduct systematic identification of outflow features \citep{2020ApJS..248...15Z, 2020ApJ...905..172X}.
\citet{2020ApJS..248...15Z} employed Support Vector Machines (SVM) to identify molecular outflows in a dark cloud complex from molecular line emission. Although SVM performs robustly in classification tasks, it requires preprocessing of the raw data cubes, i.e., manually extracting feature vectors that represent the raw data. The choice of input features are determined subjectively by visual inspection, and extracting these features from data cubes is non-trivial. Moreover, the manually extracted features discard part of the information embedded in the raw spectral cubes, which introduces uncertainty and affects the performance of classification. Convolutional neural networks (CNNs) are a powerful new approach being applied to identify structures or objects in astronomical data, such as exoplanets \citep{2018AJ....155...94S}, stellar feedback bubbles \citep{2019ApJ...880...83V, 2020ApJ...890...64X} and protostellar outflows \citep{2020ApJ...905..172X}. 
Given a labeled training set of images or spectral cube data, CNNs can be applied to efficiently identify features in large surveys.
\citet{2019ApJ...880...83V} developed a Convolutional Approach to Shell/Structure Identification, \CASI, to identify stellar feedback bubbles in 2D density slices and \co\, integrated intensity maps. \citet{2020ApJ...890...64X} and \citet{2020ApJ...905..172X} successfully extended \CASI\, to \CASItD, which is able to identify stellar feedback bubbles and protostellar outflows in position-position-velocity (PPV) molecular line spectra cubes. \citet{2020ApJ...905..172X} applied \CASItD\ to \co\ observations of the Perseus molecular cloud and identified all 60 previously visually identified outflows. Additionally, \CASItD\ found 20 new high-confidence outflows. Apart from structure identification, \CASItD\ successfully predicts hidden information in the data cube, such as the fraction of mass associated with feedback, which provides a more accurate feedback mass estimation. \citet{2020ApJ...890...64X} showed that the actual mass associated with stellar feedback in Taurus is an order of magnitude smaller than the previous visual estimates. These results illustrate the capability of CNNs to identify complex structures and infer embedded information.

In this paper, we adopt the deep learning method \CASItD\ to systemically identify protostellar outflows in the nearby molecular clouds, Ophiuchus, Taurus, Perseus and Orion. We describe \CASItD\ and the CO observations of these nearby molecular clouds in Section~\ref{Data and Method}. In Section~\ref{Results}, we present the performance of our CNN models in identifying protostellar outflows in the observational data, calculate the physical properties of outflows and discuss their impact on the clouds. We discuss the broader impact of outflows and compare them with stellar wind driven bubbles in Section~\ref{Discussion} and summarize our results and conclusions in Section~\ref{Conclusions}.

\section{Data and Method}
\label{Data and Method}

\subsection{COMPLETE Survey}
\label{COMPLETE Survey}

The \co\ J=1-0 (115.271 GHz) and \13co\ J=1-0 (110.201 GHz) lines were observed simultaneously in surveys of Ophiuchus, Taurus and Perseus between 2002 and 2005 using the 13.7 m Five College Radio Astronomy Observatory (FCRAO) Telescope \citep{2006AJ....131.2921R, 2008ApJS..177..341N}. The main beam of the antenna pattern has a FWHM of 45\arcsec\ for \co\ and 47\arcsec\ for \13co. The data are obtained on the fly (OTF), but they are resampled onto a uniform 23\arcsec\ grid \citep{2006AJ....131.2921R}.

The Ophiuchus data has root-mean-square (RMS) antenna temperatures of 0.98 K and 0.33 K for \co\ and \13co, respectively. We resampled the spectra with a lower velocity resolution of 0.125 km/s to ensure a uniform velocity resolution for all regions and to match the resolution of the training set. If the velocity resolution is too low ($>0.3$ \kms), the structure generated by stellar feedback is not distinguishable across multiple velocity channels. On the other hand, if the velocity resolution is too high ($<0.1$ \kms), the training data will hit the limitation of GPU memory. The noise levels for the new Ophiuchus \co\ and \13co\ spectra are reduced by a factor of square root of 2 to 0.69 K and 0.23 K, respectively. {The final Ophiuchus data cube has a velocity range between -0.8 and 7.5 \kms\, with 67 channels.}

The Taurus data has a RMS antenna temperature of 0.28 K for \co\ and 0.125 K for \13co. There are 80 and 76 channels with 0.26 and 0.27~\kms\, spacing for \co\ and \13co, respectively. {The velocity range of the Taurus data spans -5.1 to 14.9 \kms.}

The Perseus \co\ and \13co\ data have a RMS antenna temperature of 0.25 K and 0.2 K, respectively. We resampled the spectra with a lower velocity resolution of 0.125 km/s to ensure a uniform velocity resolution for all regions and to match the resolution of the training set. The noise levels for the new \co\ and \13co\ spectra is reduced by a factor of square root of 2 to 0.17 K and 0.14 K, respectively. {The final Perseus data cube has a velocity range between -2.0 and 15.0 \kms\, with 137 channels.}

\subsection{NRO45 Orion Survey}
\label{NRO45 Orion Survey}

The \co\ J=1-0 and \13co\ J=1-0 observations of Orion A were carried out from 2007 to 2017 by the Nobeyama Radio Observatory 45 m telescope (NRO 45m), using two different receivers \citep{2015ApJS..217....7S,2015ApJS..221...31S,2019PASJ...71S...9I,2019PASJ...71S...3N}. The two maps were calibrated to the same intensity scale and combined on a common grid with a pixel scale of 7\arcsec.5{, which corresponds to an effective angular resolution of 22\arcsec. \citet{2019PASJ...71S...9I} smoothed the combined map to a velocity resolution of 0.22 km/s, and further converted the intensity to the main beam temperature scale that is benchmarked with data from FCRAO and IRAM 30m.} The final sensitivity for the \co\ map is 0.35 K, while the sensitivity for the \13co\ map is 0.40 K. {The velocity range of the Orion data spans -2.9 to 19.3 \kms.} More detailed description about the data can be found in \citet{2018ApJS..236...25K} {and \citet{2019PASJ...71S...9I}}.

\subsection{YSO Catalogue}
\label{YSO Catalogue}

To validate our outflow identifications, we compare our feedback maps with the observed distributions of YSOs. We use the YSO catalog for Ophiuchus and Perseus from SESNA \citep[Spitzer Extended Solar Neighborhood Archive,][]{Gutermuth.in.prep} used by \citet{2020ApJ...896...60P}. SESNA uses an updated implementation of the data treatment, source catalog construction, and YSO identification and classification processes on Spitzer surveys \citep{2009ApJS..184...18G}. SESNA classified YSOs into four groups: deeply embedded protostars, Class I YSOs, Class II YSOs, and transition disks. For further analysis, we combine the former two groups as ``Younger YSOs'' and merge the latter two groups as ``Older YSOs''.

We adopt the YSO catalog from \citet{2010ApJS..186..259R} for Taurus, which contains 215 previously identified YSOs and 148 newly identified YSOs. \citet{2010ApJS..186..259R} adopted observations from IRAC and MIPS between 2005 and 2007. \citet{2010ApJS..186..259R} applied color-magnitude criteria to select the YSO candidates and classified these YSO candidates into four different classes: Class I, II, III and Flat. For further analysis, we combine YSOs in Class I and Flat into a single category of ``Younger YSOs'' and merge YSOs in Class II and III into a single category of ``Older YSOs''.

We adopt the YSO catalog from \citet{2012AJ....144..192M} for Orion, which contains a total of 3479 dusty YSOs from IRAC and MIPS observations. \citet{2012AJ....144..192M} classified these YSOs into three categories, which includes 2991 young stars with disks, 428 protostars, and 50 faint candidate protostars. For further analysis, we designate protostars as ``Younger YSOs'' and young stars with disks as ``Older YSOs''.

YSOs in the group of ``Younger YSOs'' have an average age smaller than 1 Myr, while YSOs in the group of ``Older YSOs'' have an average age around 1-3 Myrs \citep{2014prpl.conf..195D}. However, it is worth noting that truly younger YSOs and older YSOs may be misclassified based on their orientation with respect to the line of sight \citep{2012ApJ...753...98O}, which means the actual age of an individual YSO in the group of ``Younger YSOs'' is not necessarily smaller than that of an object in the group of ``Older YSOs''. 

Finally, we note that not all the surveys completely cover the area of observed molecular emission. We focus our analysis on the areas with YSO coverage and indicate the boundaries of the surveys on the maps.


\subsection{Method} 
\label{Method}

\subsubsection{\CASItD}
\label{CASItD}

We adopt the previously trained \CASItD\, model from \citet{2020ApJ...905..172X} to identify protostellar outflows in \co. \CASItD\, is an encoder-decoder based convolutional neural network combining both residual networks \citep{He2016} and a ``U-net'' \citep{Ronneberger2015}. \CASItD\, is trained on synthetic \co\ data, which models forming stars that launch protostellar outflows. We adopt the same training set and ranges of magnetohydrodynamic (MHD) model properties as described in {\citet[][Table~1]{2020ApJ...905..172X}}. We adopt three different \co\ abundances and two different cloud kinetic temperatures when conducting synthetic observations as described in {\citet[][Table~2]{2020ApJ...905..172X}}. \CASItD\ takes \co\ data cubes as input and predicts the position of outflows on the voxel level. \citet{2020ApJ...905..172X} trained two \CASItD\ models to identify protostellar outflows. One model, model ME1, is trained to predict the {\it \co\ emission} that is associated with outflows. The other model, model MF, is trained to predict {\it the fraction of the mass} that comes from stellar feedback. A more detailed description of how we generate training data for these two models can be found in {\citet{2020ApJ...905..172X}}. 

\subsubsection{Data Preprocessing}

\begin{figure}[hbt!]
\centering
\includegraphics[width=0.98\linewidth]{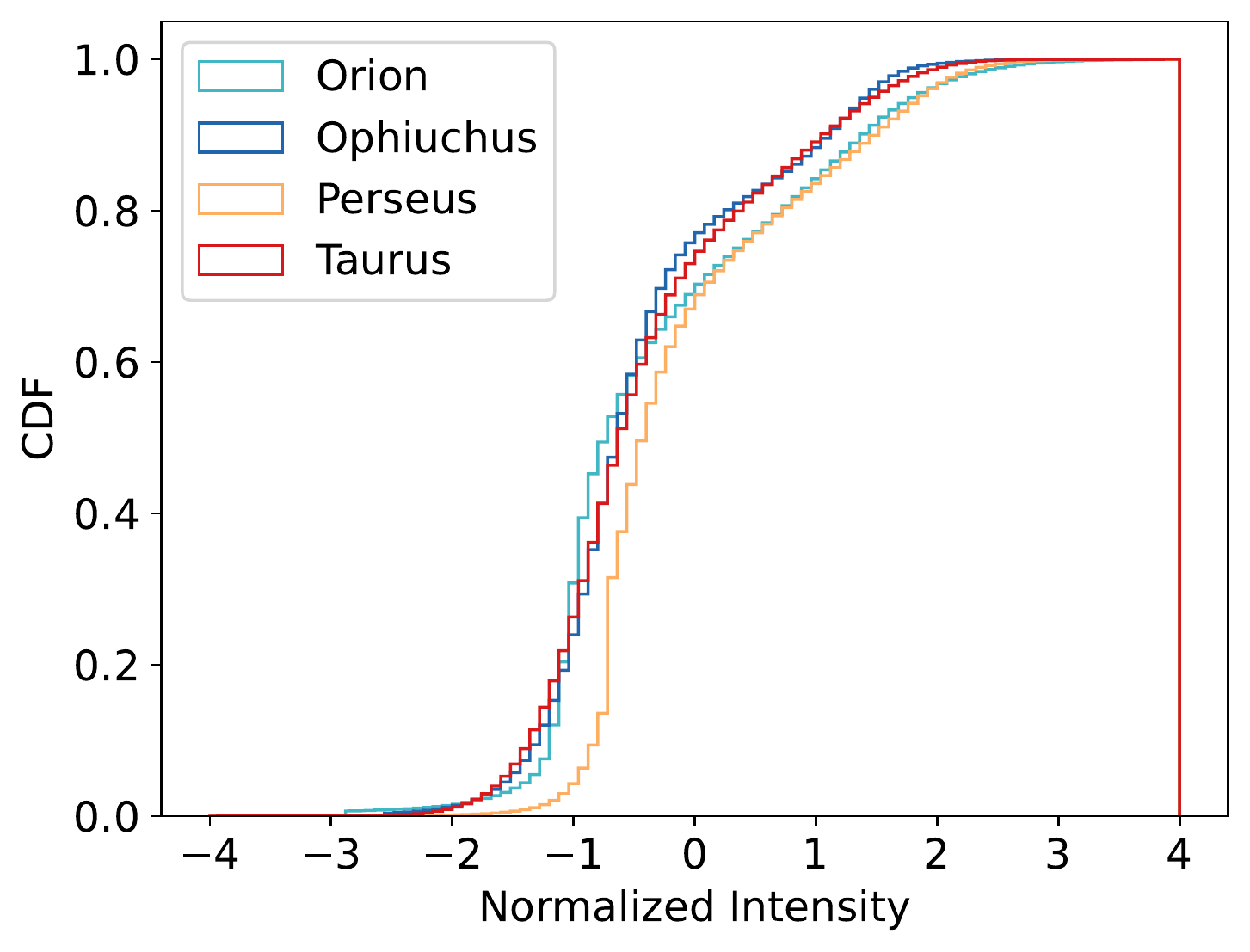}
\caption{Cumulative distribution function (CDF) of the normalized emission for the four clouds.}
\label{fig.stat-T-norm-hist}
\end{figure}

Before we apply our \CASItD\ models to the observational data, we apply the same preprocessing steps that we adopt for the training data. Due to the relatively large dynamic range of the \co\ emission, we take the logarithm of the emission and then normalize the values by subtracting the mean and dividing by the standard deviation of the full map. Figure~\ref{fig.stat-T-norm-hist} illustrates the cumulative distribution of normalized emission for the four clouds. Although the \co\ emission varies significantly between clouds, after preprocessing, the normalized emission of the four clouds is similar. This allows \CASItD\ to perform stably across a variety of conditions. 

\citet{2020ApJ...905..172X} examined the performance of  \CASItD\ on conditions that are not included in the training set, such as different kinetic temperatures, \co~abundances, beam sizes and noise levels. {\CASItD\ is able to successfully identify outflows across a variety of physical and observational conditions.} See Appendix C in \citet{2020ApJ...905..172X} for more details. These tests suggest that the \CASItD\ identifications are relatively robust across the range of conditions expected to be characteristic of the observed clouds.

After preprocessing the observational data, we adopt the same strategy as \citet{2020ApJ...890...64X} to carry out the full \co\ map prediction. We crop the full \co\ map into a stack of three-dimensional chunks. Each chunk has a size of $64\times 64\times 32$ (position-position-velocity). To reduce the bias due to the position of outflows, each chunk has at least 84\% volume overlap with adjacent chunks. To construct the full-map prediction, we combine the predictions for the individual chunks and adopt the maximum predicted value for each voxel.

\subsubsection{Mass Calculations}
\label{Mass Calculations}

We follow the same strategy as \citet{2010ApJ...715.1170A} to calculate the outflow mass by combining both \co\ and \13co\ data. If there is distinct \13co\ emission at the corresponding position, we use \13co\ to calculate the outflow mass. We assume the \13co\ emission line is optically thin and the \co\ emission line is optical thick. We adopt an excitation temperature of 25 K or the \co\ peak temperature, whichever is higher, to calculate the mass \citep{2010ApJ...715.1170A,2012MNRAS.425.2641N,2015ApJS..219...20L,2020ApJ...896...11F}. If there is no distinct \13co\ emission at the corresponding position, we assume the \co\ emission line is optically thin to derive the mass. Under the assumption of local thermodynamic equillibrium (LTE), the mass estimation scales linearly with the excitation temperature. From previous feedback mass estimates, the choice of excitation temperature ranges from 10 K to 50 K. This could potentially introduce a factor of two uncertainty in the mass estimation. We take 62 as the abundance ratio between \co\ and \13co\ and 10$^{-4}$ as the abundance ratio between \co\ and \h2 \citep{2010ApJ...715.1170A,2020ApJ...896...11F}. {We verify that the \13co\ emission is generally optically thin for all four clouds. For example, Orion  has the largest cloud mass and highest column density among the four clouds. The optical depth of \13co\ in Orion is less than 1 for 99.4\% of the pixels \citep[see also][]{2018ApJS..236...25K}.
Only 0.6\% of the pixels have $\tau_{13} >$ 1, and these are mostly in the OMC-2/3 and L1641-N regions. The maximum value of $\tau_{13}$ is 7.8. 
}

\section{Results}
\label{Results}

\subsection{Outflows Identified in the Full Map}
\label{Outflows Identified in the Full Map}

In this section, we present the prediction by models ME1 and MF on the four star-forming clouds (Ophiuchus, Taurus, Perseus and Orion). Most molecular clouds have a global velocity gradient, which indicates there is no one unique central velocity of a cloud. To better visualize the blue-shifted and red-shifted lobes of outflows, we remove the large scale gradient to make local outflow motions clearer, i.e., shifting the velocity zero point to the flux-weighted central velocity of the spectrum for each pixel. {We calculate the flux-weighted central velocity from the 1$^{st}$ moment maps. In order to reduce the velocity fluctuations on small scales, we apply a Gaussian kernel with a FWHM that corresponds to 0.1 pc to convolve the 1$^{st}$ moment maps.} For each pixel, we show the integrated prediction over the channels that have absolute velocities greater than the central velocity by the specified thresholds.

\subsubsection{Ophiuchus}
\label{Ophiuchus}

\begin{figure*}[hbt!]
\centering
\includegraphics[width=0.98\linewidth]{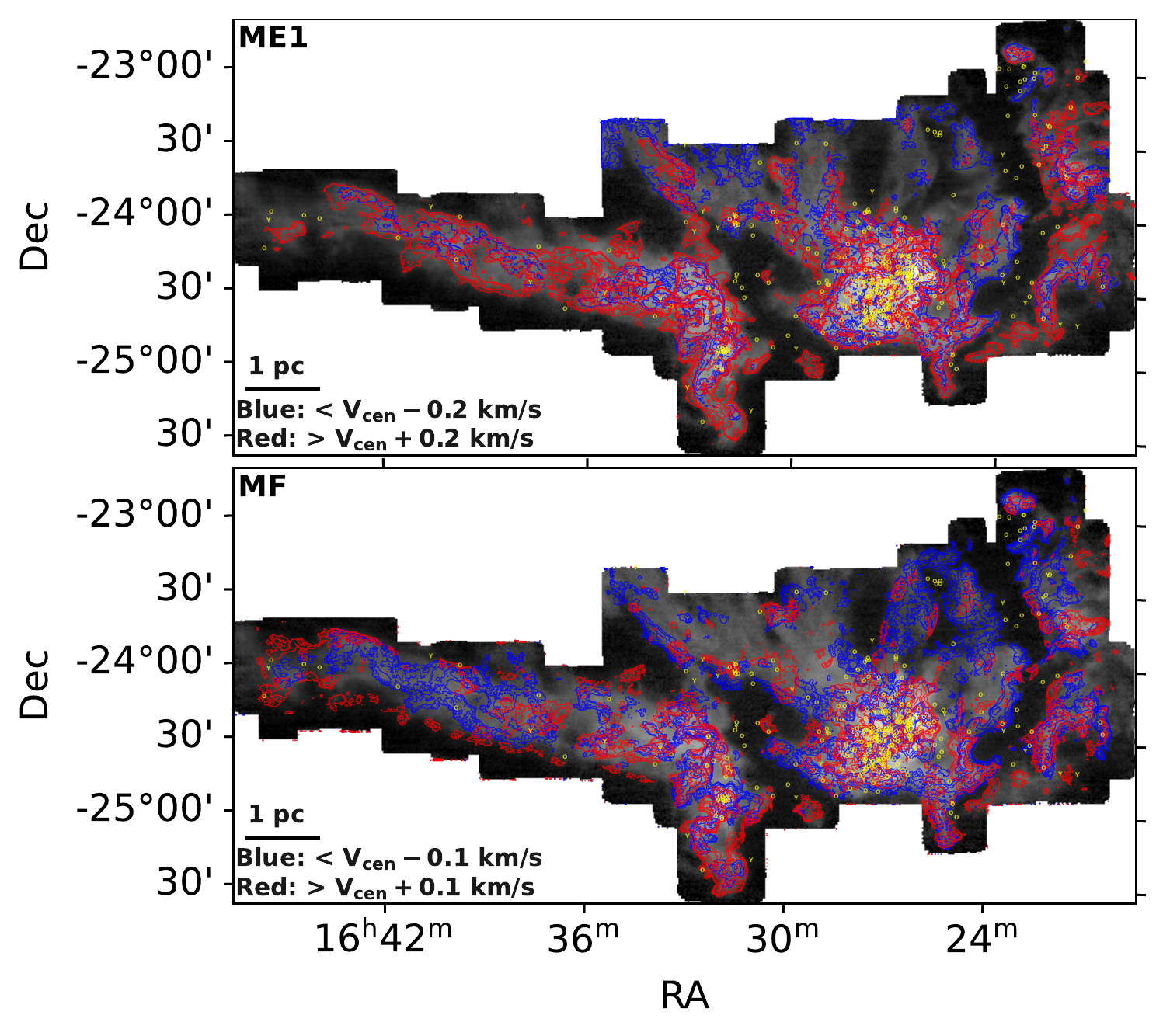}
\caption{Intensity of \co\ (1-0) integrated over all velocity channels for Ophiuchus, overlaid with the model ME1 prediction (upper panel in red and blue contours) and with the model MF prediction (lower panel in red and blue contours). Red contours indicate the integrated prediction over the channels that have velocities greater than $V_{\rm cen}+0.4$ \kms\ for model ME1 and $V_{\rm cen}+0.2$ \kms\ for model MF. Blue contours indicate the integrated prediction over the channels that have velocities smaller than $V_{\rm cen}-0.4$ \kms\ for model ME1 and $V_{\rm cen}-0.2$ \kms\ for model MF. ``Y'' and ``O'' indicates the location of YSOs, as described in Section~\ref{YSO Catalogue}.  The contours start at 25th percentile of the sorted pixel values and end at 99.7th percentile of the sorted pixel values of the data, with 6 levels evenly spaced. It is worth noting that the absolute values of the contour levels for models ME1 and MF are different. {The contour levels for the model ME1 prediction start at 2.8 \kkms, for the blue-shifted lobes and 2.2 \kkms, for the red-shifted lobes, and end at 19.3 \kkms, and 15.3 \kkms, for the blue- and red-shifted lobes, respectively. The contour levels for the model MF prediction start at 0.80 \kkms, (blue) and 0.96 \kkms, (red), and end at 7.2 \kkms, (blue) and 8.6 \kkms, (red).} The method of plotting contours is the same for Figure~\ref{fig.pred-taurus-lm-ME1}, \ref{fig.pred-taurus-lm-MF}, \ref{fig.pred-perseus-lm-ME1} and \ref{fig.pred-orion-lm-ME1}.}
\label{fig.pred-oph-lm-ME1}
\end{figure*}

Figure~\ref{fig.pred-oph-lm-ME1} shows ME1 and MF model predictions for Ophiuchus. Both models identify a large amount of outflow activity in the region of the large, central star cluster. In contrast, the models identify much less outflow activity just north of the cluster, which has only a few YSOs.

\begin{figure*}[hbt!]
\centering
\includegraphics[width=0.98\linewidth]{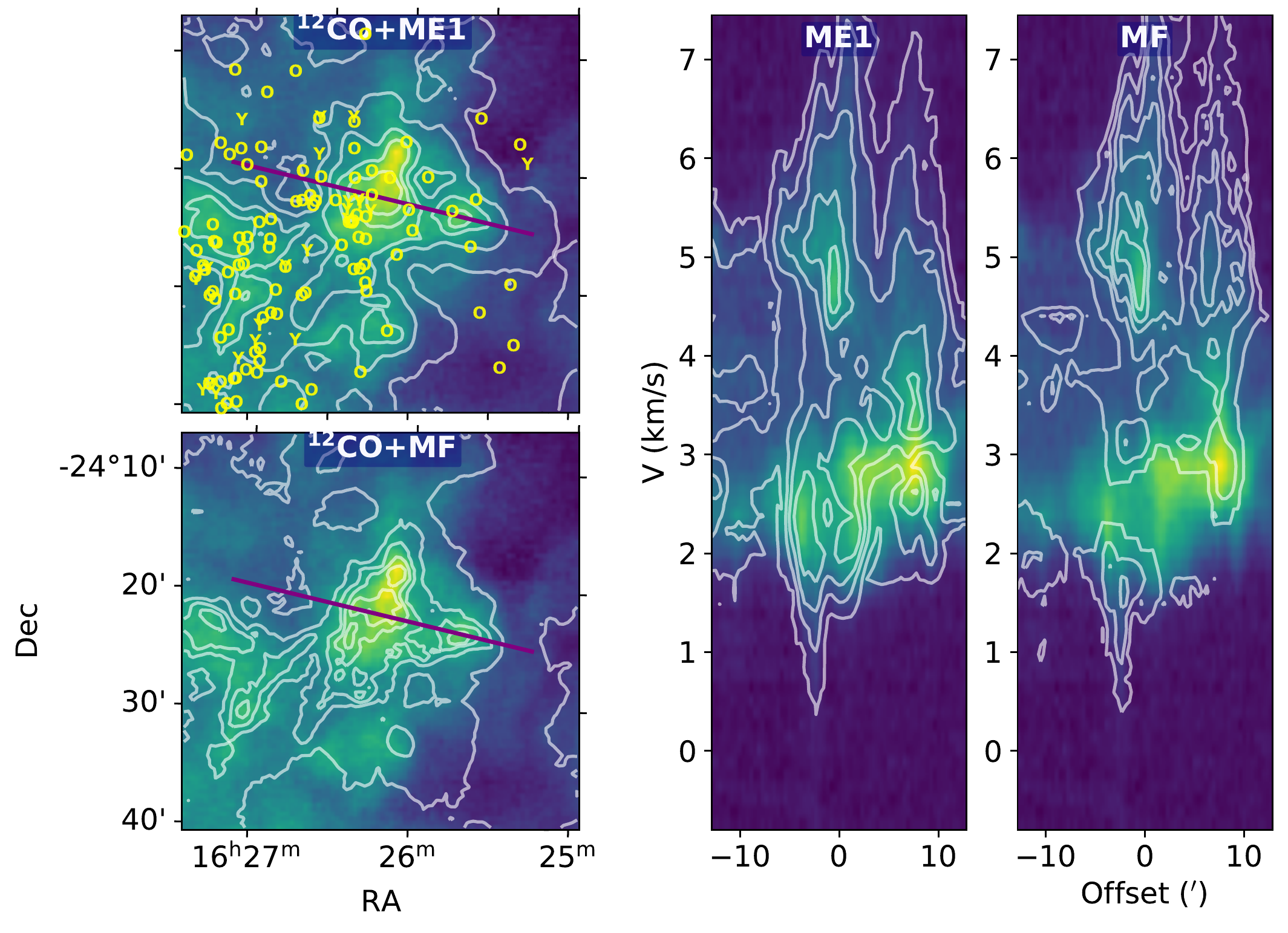}
\caption{Position-velocity diagram of \co\ emission toward the Ophiuchus L1688 region. Left panel: integrated intensity of \co\ over the the full velocity range (from -0.8 km/s to 7.5 km/s) overlaid with the model ME1 and MF predictions in white contours. Letters ``Y'' and ``O'' mark YSO positions, as described in Section~\ref{YSO Catalogue}. The purple line illustrates the cut direction of the position-velocity diagram. Middle and right panel: position-velocity diagram of \co\ emission overlaid with the model ME1 and MF predictions in white contours. }
\label{fig.pv-map-oph-goodcase}
\end{figure*}

Figure~\ref{fig.pv-map-oph-goodcase} shows the predicted outflow activity by models ME1 and MF toward the young star cluster region Lynds 1688 
in Ophiuchus. The L1688 region has over 120 young stars and a large number of interacting outflows. Model ME1, which performs similarly to human visual identification, identifies almost everything as feedback in this region. In contrast, Model MF provides an approach to disentangle the outflow emission from the ambient cloud emission. Quantitively, model ME1 predicts 81\% of the mass in this region as outflows. However, model MF predicts that only 12\% of the total mass is associated with outflows. Figure~\ref{fig.pv-map-oph-goodcase} shows that both models ME1 and MF are able to identify coherent high velocity structures in the position-velocity diagram. 


It is challenging to separate outflow emission from the cloud emission in \co~(1-0) ``by eye," since the cloud emission dominates the \co~(1-0) emission. Consequently, even an expert astronomer can only separate the high-velocity component from the host cloud emission but cannot identify the outflow morphology near the rest frame velocity of the cloud.

\subsubsection{Taurus}
\label{Taurus}

\begin{figure*}[hbt!]
\centering
\includegraphics[width=0.98\linewidth]{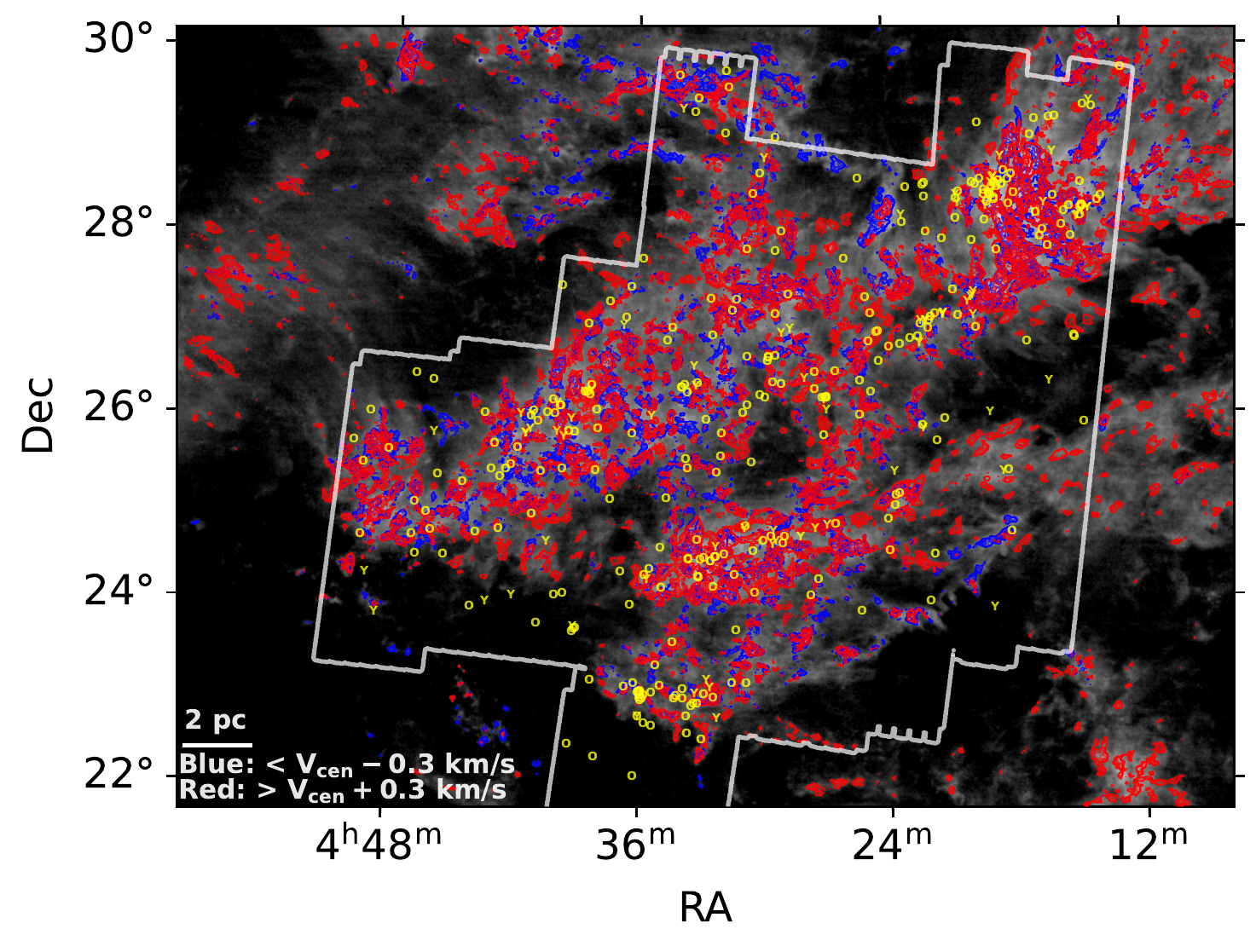}
\caption{Intensity of \co\ (1-0) integrated over all velocity channels for Taurus, overlaid with the model ME1 prediction (red and blue contours). Letters ``Y'' and ``O'' mark YSO positions, as described in Section~\ref{YSO Catalogue}. The grey line encloses the Spitzer coverage, where YSOs are identified \citep{2010ApJS..186..259R}. {The contours start at the 25th percentile of the sorted pixel values (0.47 \kkms, for blue contours and 0.49 \kkms, for red contours) and end at the 99.7th percentile of the sorted pixel values of the data (2.9 \kkms, for blue contours and 3.4 for red contours), with 6 levels evenly spaced. Note that the absolute values of the contour levels for models ME1 and MF are different.}}
\label{fig.pred-taurus-lm-ME1}
\end{figure*}

\begin{figure*}[hbt!]
\centering
\includegraphics[width=0.98\linewidth]{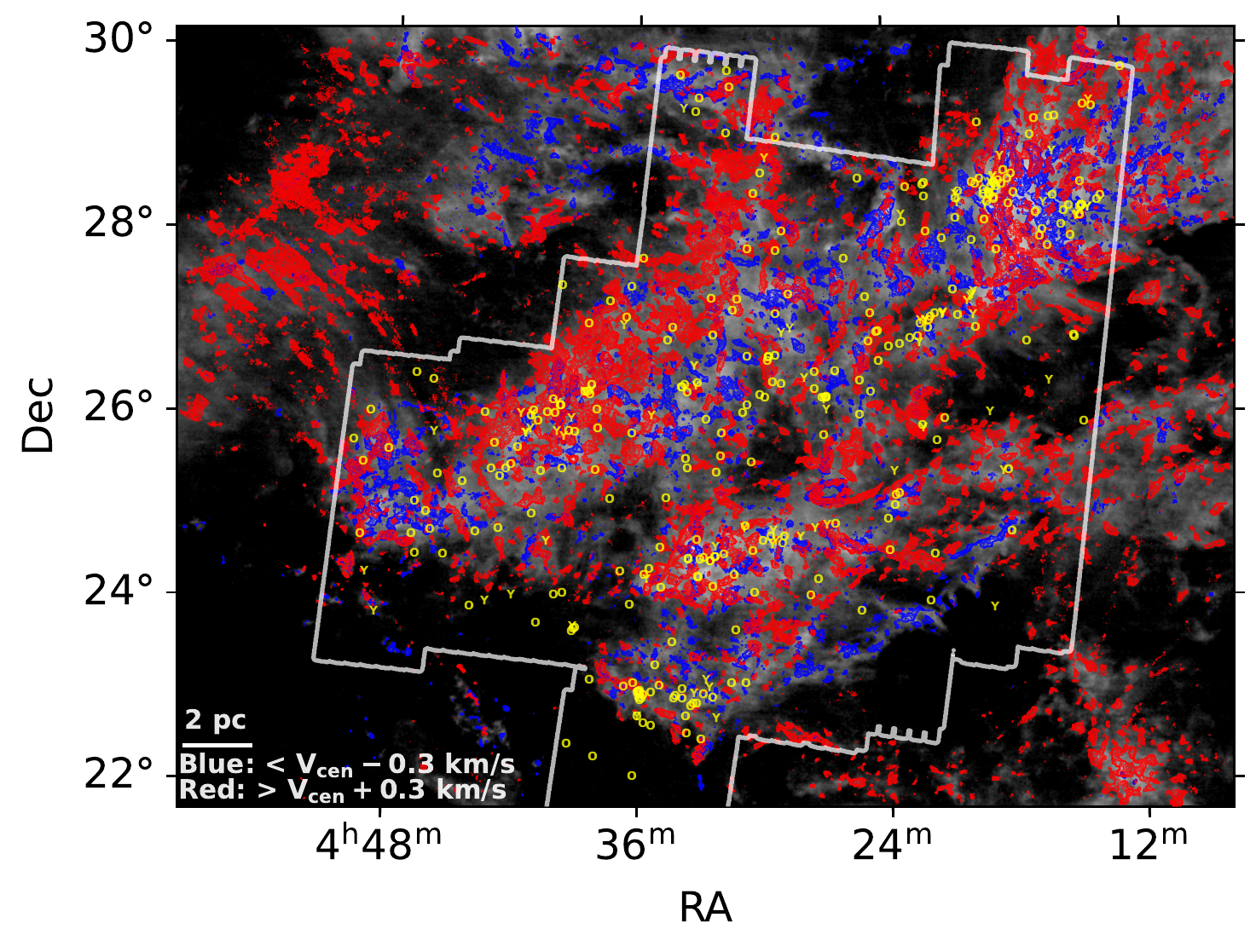}
\caption{Intensity of \co\ (1-0) integrated over all velocity channels for Taurus, overlaid with the model MF prediction (red and blue contours). Letters ``Y'' and ``O'' mark YSO positions, as described in Section~\ref{YSO Catalogue}. The grey line encloses the Spitzer coverage, where YSOs are identified \citep{2010ApJS..186..259R}. {The contours start at the 25th percentile of the sorted pixel values (0.27 \kkms, for blue contours and 0.27 \kkms, for red contours) and end at the 99.7th percentile of the sorted pixel values of the data (1.7 \kkms, for blue contours and 1.5 for red contours), with 6 levels evenly spaced.}}
\label{fig.pred-taurus-lm-MF}
\end{figure*}

Figure~\ref{fig.pred-taurus-lm-ME1} and \ref{fig.pred-taurus-lm-MF} show the ME1 and MF model predictions for Taurus. The structures in the outflow prediction maps are more discrete than those in Ophiuchus. One possible reason is that YSOs in Taurus are more sparsely distributed than those in Ophiuchus. Although the total number of YSOs in Taurus is slightly larger, they are less clustered. In Ophiuchus, the outflows driven by YSOs are more likely to interact and overlap with each other. While in Taurus, the outflows are more isolated. 

\begin{figure*}[hbt!]
\centering
\includegraphics[width=0.98\linewidth]{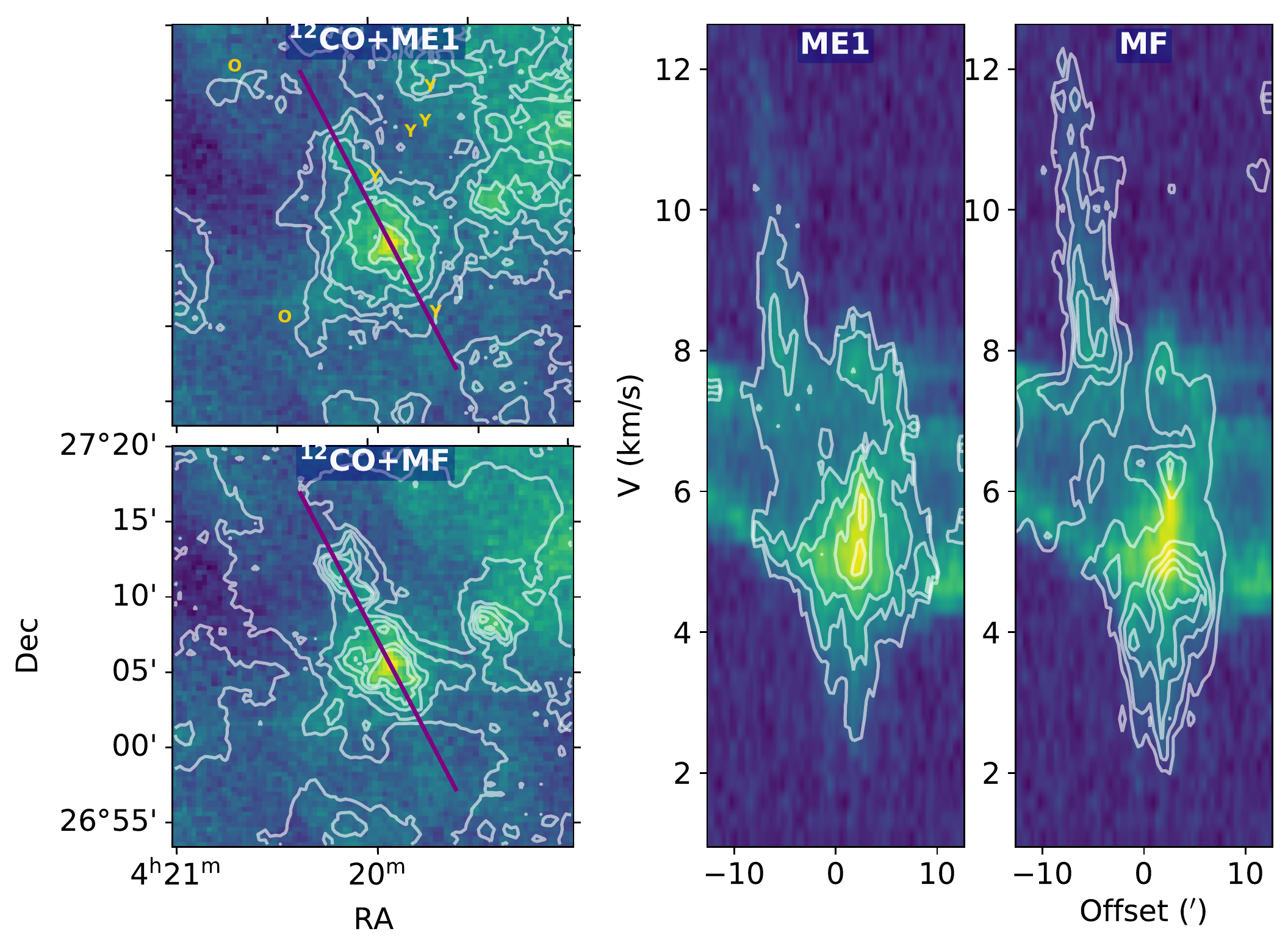}
\caption{Position-velocity diagram of \co\ emission toward a previously identified outflow, TMO\_06 \citep{2015ApJS..219...20L}, in Taurus. Left panel: integrated intensity of \co\ over the the full velocity range (from -1.5 km/s to 13.4 km/s) overlaid with the model ME1 and MF predictions in white contours. Letters ``Y'' and ``O'' mark YSO positions, as described in Section~\ref{YSO Catalogue}. The purple line illustrates the cut direction of the position-velocity diagram. Middle and right panel: position-velocity diagram of \co\ emission overlaid with the model ME1 and MF predictions in white contours. }
\label{fig.pv-map-taurus-goodcase}
\end{figure*}

Figure~\ref{fig.pv-map-taurus-goodcase} shows the outflow activity predicted by models ME1 and MF toward a previously identified outflow, TMO\_06 \citep{2015ApJS..219...20L}, in Taurus. Several young YSOs are located around this outflow. Both models are able to identify the blueshifted and redshifted lobes of this outflow. However, the model MF prediction is more extended in the position-velocity diagram. In high velocity channels, the \co\ emission produced by outflows is usually faint while the fraction of outflow mass is high. Model ME1 only identifies the location of outflows and gives stronger emission a higher weight. In this subregion, model ME1 predicts that 43\% of the mass in this region is associated with outflows. However, model MF predicts that only 8\% of the total mass is outflows gas.

\subsubsection{Perseus}
\label{Perseus}

\begin{figure*}[hbt!]
\centering
\includegraphics[width=0.98\linewidth]{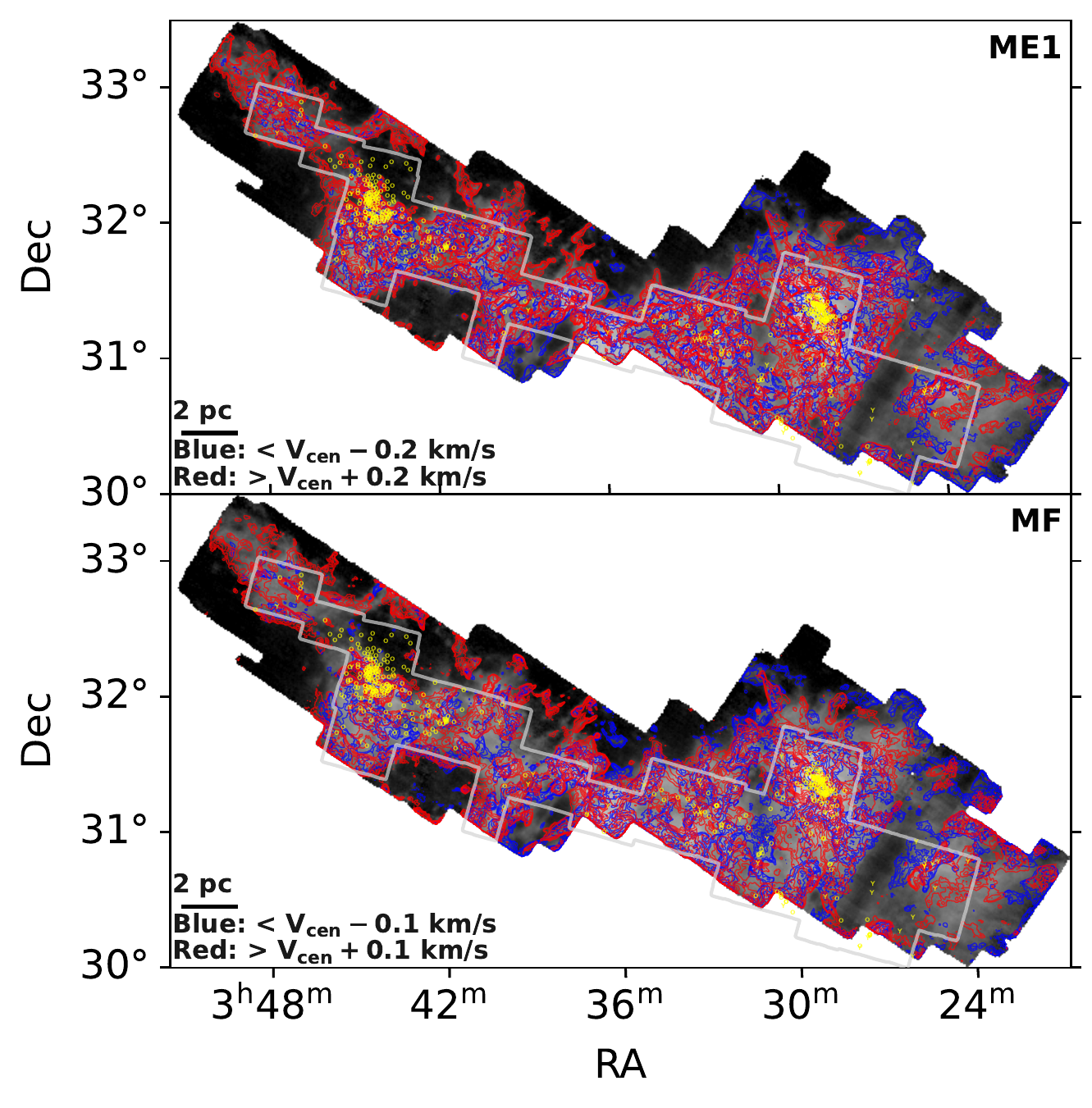}
\caption{Intensity of \co\ (1-0) integrated over all velocity channels for Perseus, overlaid with the model ME1 prediction (upper panel in red and blue contours) and with the model MF prediction (lower panel in red and blue contours). Letters ``Y'' and ``O'' mark YSO positions, as described in Section~\ref{YSO Catalogue}. The grey line encloses the Spitzer coverage, where YSOs are identified \citep{Gutermuth.in.prep}. {The contour levels for the model ME1 prediction start at 2.5 \kkms\, (blue) and 2.1 \kkms\, (red), and end at 17 \kkms\, (blue) and 15 \kkms\, (red). The contour levels for the model MF prediction start at 0.96 \kkms\, (blue) and 0.90 \kkms\, (red), and end at 8.6 \kkms\, (blue) and 8.0 \kkms\, (red).}}
\label{fig.pred-perseus-lm-ME1}
\end{figure*}

\citet{2020ApJ...905..172X} followed up previous visually identified outflow targets and validated the model performance. Here we extend that analysis by carrying out a ``blind search'' where we analyze the full cloud. Figure~\ref{fig.pred-perseus-lm-ME1} shows the full map predictions for both models. Both model ME1 and MF predictions are more concentrated towards star clusters. The model ME1 prediction is more spatially extended than that by model MF. This might be caused by a strong emission region with a low outflow mass fraction, where model ME1 counts the entire voxel as feedback while model MF recognizes that much of this emission is cloud contamination and excludes it.

\subsubsection{Orion}
\label{Orion}

\begin{figure*}[hbt!]
\centering
\includegraphics[width=0.98\linewidth]{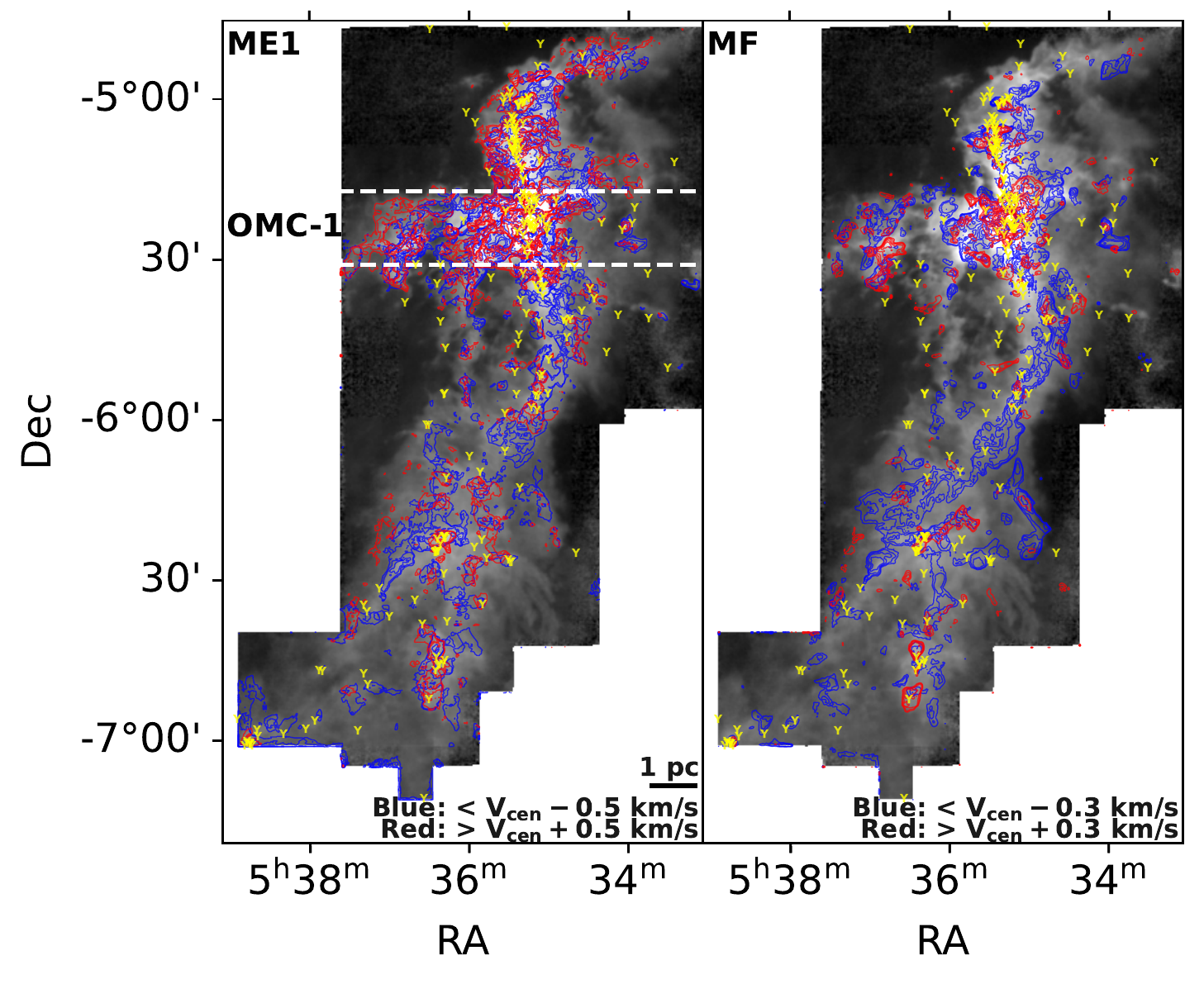}
\caption{Intensity of \co\ integrated over all velocity channels for Orion, overlaid with the model ME1 prediction (left panel in red and blue contours) and with the model MF prediction (right panel in red and blue contours). Letters ``Y'' mark young YSO positions, as described in Section~\ref{YSO Catalogue}. {The contour levels for the model ME1 prediction start at 2.1 \kkms\, (blue) and 2.6 \kkms\, (red), and end at 14 \kkms\, (blue) and 18 \kkms\, (red). The contours levels for the model MF prediction start at 0.48 \kkms\, (blue) and 0.55 \kkms\, (red), and end at 4.3 \kkms\, (blue) and 4.9 \kkms\, (red).}}
\label{fig.pred-orion-lm-ME1}
\end{figure*}

Figure~\ref{fig.pred-orion-lm-ME1} shows the model ME1 and MF predictions for Orion. Due to the large number of old evolutionary stage YSOs in Orion, we only show the younger YSOs to reduce confusion. We show the position of all Orion YSOs in Figure~\ref{fig.yso-orion-lm-all} in Appendix~\ref{YSOs in Orion}. The prediction by both models covers most of the emission in Orion, which indicates outflows exist everywhere. This is consistent with the extremely dense distribution of YSOs in Orion. 

\begin{figure*}[hbt!]
\centering
\includegraphics[width=0.98\linewidth]{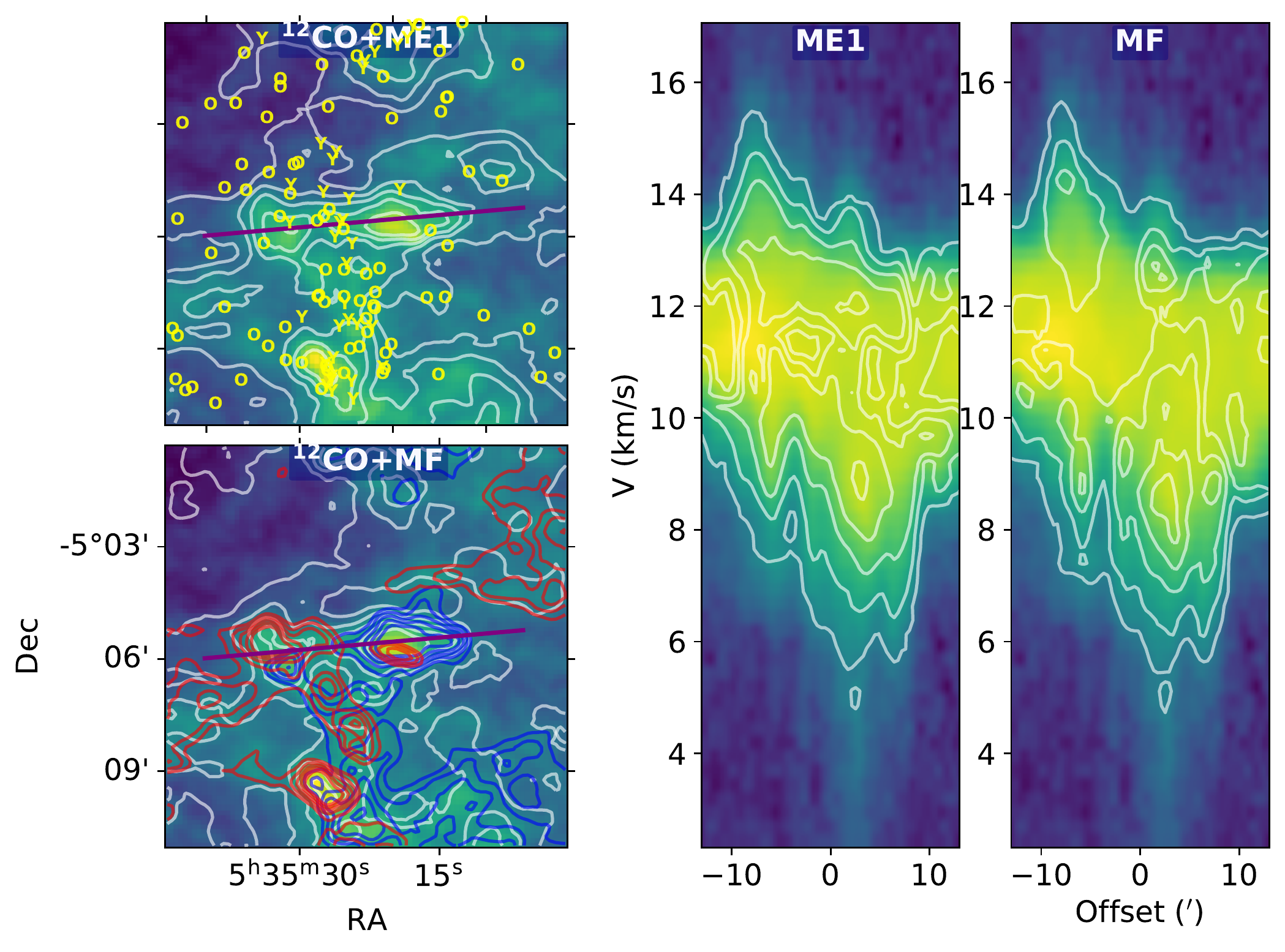}
\caption{Position-velocity diagram of \co\ emission toward a previously identified outflow, Outflow No. 7 \citep{2019PASJ...71S...8T}, in Orion. Left panel: integrated intensity of \co\ over the the full velocity range (from 2.3 km/s to 17.1 km/s) overlaid with the model ME1 and MF predictions in white contours. The blue and red contours indicate the high velocity components, integrated over the raw emission with velocity ranges of 2.3-9.5 \kms and 12.3-17.0 \kms, respectively. Letters ``Y'' and ``O'' mark YSO positions, as described in Section~\ref{YSO Catalogue}. The purple line illustrates the cut direction of the position-velocity diagram. Middle and right panel: position-velocity diagram of \co\ emission overlaid with the model ME1 and MF predictions in white contours. }
\label{fig.pv-map-orion-goodcase}
\end{figure*}

Figure~\ref{fig.pv-map-orion-goodcase} shows the predicted outflow activity by models ME1 and MF toward a previously identified outflow, Outflow No. 7 \citep{2019PASJ...71S...8T}, in Orion. Both models similarly highlight the coherent high-velocity structures. There are several young YSOs located around this outflow, which could be the driving source. For reference, we illustrate the high-velocity components in blue and red contours, which are considered to be outflows by \citet{2019PASJ...71S...8T}. The mass of these high-velocity components is 268 \msun. Model MF predicts the outflow mass to be 308 \msun, which is similar to previous estimates. In this subregion, model ME1 predicts 61\% of the mass in this region is associated with outflows. However, model MF predicts that only 13\% of the total mass is outflow gas.


\subsection{Physical Properties of Outflows}
\label{Physical Properties of Outflows}

In this section, we study the physical properties of the outflows we identified and analyze the correlation between the physical quantities of outflows and the number of YSOs in the clouds. This analysis help us evaluate the robustness of the outflow identification by \CASItD. If the outflows are accurately identified by \CASItD, we should expect a linear correlation between the outflow mass and the number of YSOs. This is exactly what we find in this section and Section~\ref{Correlation Between the Outflow Properties and the Number of YSOs}. Because our analysis aims to correlate the physical properties of outflows with the number of YSOs, we only calculate mass, energy and momentum of outflows in the source catalog covered area.

\begin{figure*}[hbt!]
\centering
\includegraphics[width=0.98\linewidth]{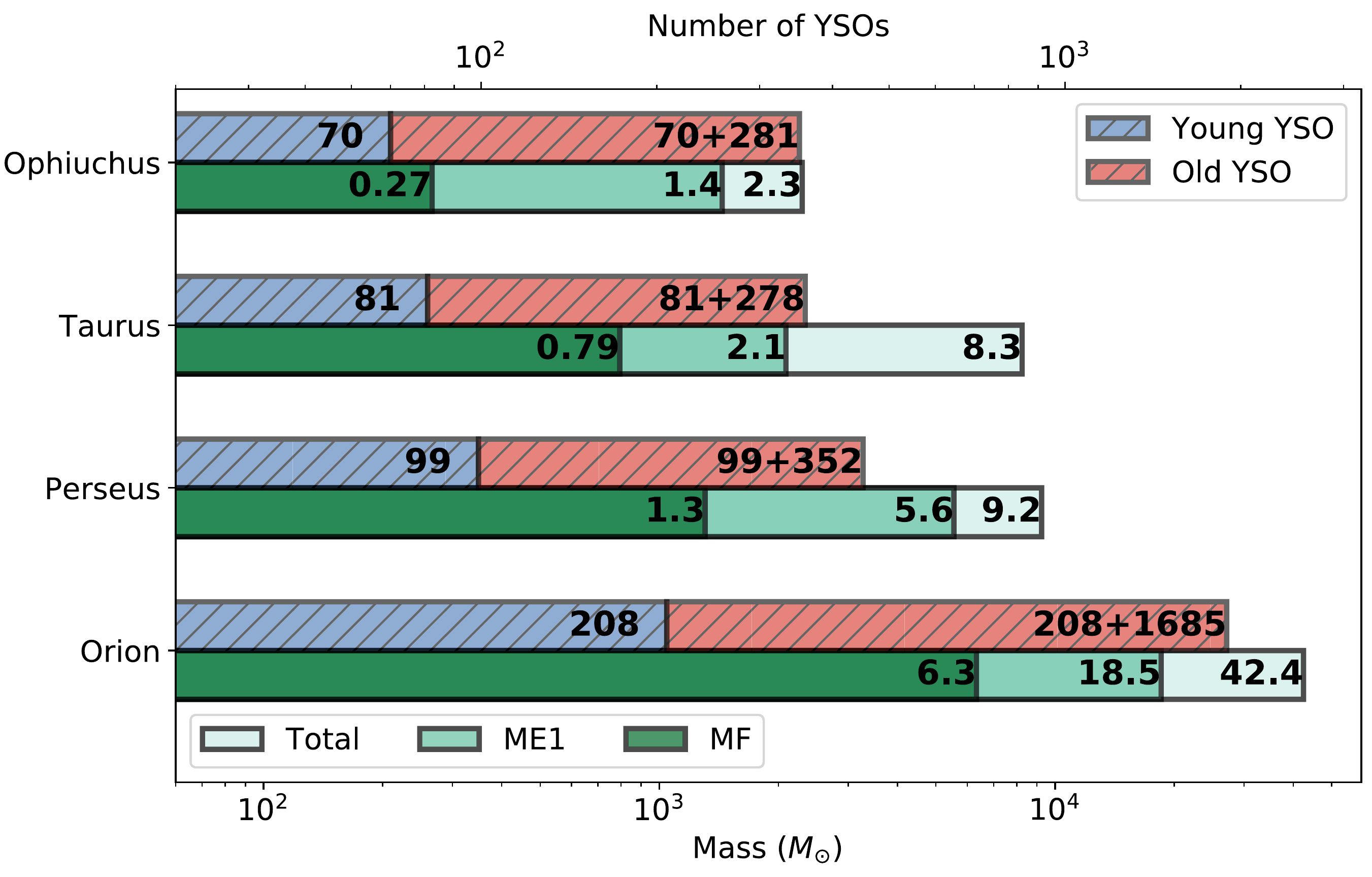}
\caption{Outflow mass (solid green colors) and the number of YSOs (hatched blue and red) in the four molecular clouds. Numbers in the hatched bars indicate the number of YSOs. Numbers in the unhatched bars indicate the gas mass in units of $10^{3}$ \msun. The total mass indicates the mass of the host cloud within the area studied. }
\label{fig.stat-mass-all}
\end{figure*}

We follow the method in Section~\ref{Mass Calculations} to calculate the mass of outflows. We adopt the model MF estimates for the fiducial values.  Figure~\ref{fig.stat-mass-all} shows the mass estimates for the outflows in the four molecular clouds. When there are more YSOs in the region, the predicted outflow mass is higher. In Ophiuchus, on average, each YSO (including both young and old) contributes 0.43 \msun\ outflow mass to the host cloud. While the values are 1.7 \msun, 2.8 \msun, and 3.4 \msun\ for Taurus, Perseus and Orion, respectively. On average, in each of the four clouds, the total mass associated with feedback is about 10\% of the mass of the host cloud within the area studied.   

We define the 1D (line-of-sight, LOS) momentum as the sum of the gas mass in each channel multiplied by the channel velocity, where we have shifted the mean cloud velocity to zero. To better quantify LOS momentum, we subtract the central velocity along each sightline to reduce the effect of large velocity gradients across the entire cloud, as described in Section~\ref{Outflows Identified in the Full Map}. Subtracting the velocity gradient provides a better estimate of LOS momentum, which reduces the impact of large-scale motions on our estimates. For comparison, we also calculate the LOS moment without subtracting the velocity gradient. The LOS momenta of the outflows and the host clouds are calculated by equations: 
\begin{equation}
\label{momentum-eq1}
P(VGsub) =\sum_{i,j,k}M_{{\rm CO},i,j,k}(v_{i,j,k}-\overline{v_{i,j}}) ,
\end{equation}

\begin{equation}
\label{momentum-eq2}
P(VGnonsub) =\sum_{i,j,k}M_{{\rm CO},i,j,k}(v_{i,j,k}-\overline{v_{\rm global}}) .
\end{equation}
Similarly, we define the LOS kinetic energy as
\begin{equation}
\label{energy-eq1}
E(VGsub) =\sum_{i,j,k}\frac{1}{2}M_{{\rm CO},i,j,k}(v_{i,j,k}-\overline{v_{i,j}})^{2} ,
\end{equation}

\begin{equation}
\label{energy-eq2}
E(VGnonsub) =\sum_{i,j,k}\frac{1}{2}M_{{\rm CO},i,j,k}(v_{i,j,k}-\overline{v_{\rm global}})^{2} .
\end{equation}

\begin{figure*}[hbt!]
\centering
\includegraphics[width=0.98\linewidth]{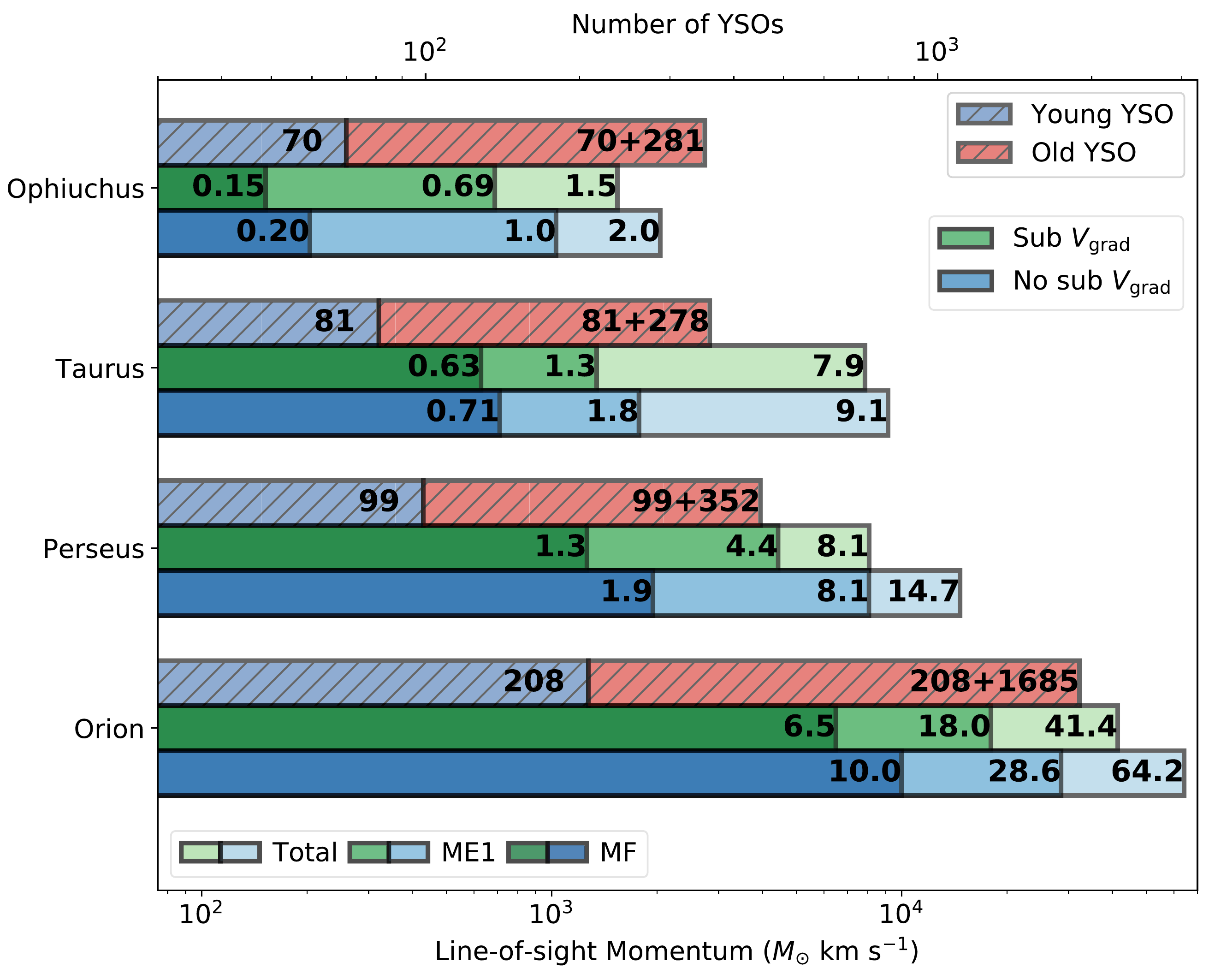}
\caption{Outflow line-of-sight momentum (solid colors) and the number of YSOs (hatched colors ) in the four molecular clouds. Numbers in the hatched bars indicate the number of YSOs. Numbers in the unhatched bars indicate the gas momentum in units of $10^{3}$ \msun\ \kms. The total momentum indicates the momentum of the host cloud within the area studied.}
\label{fig.stat-momentum-all}
\end{figure*}

Figure~\ref{fig.stat-momentum-all} shows the LOS outflow momentum estimates for the four molecular clouds. We list two sets of momentum estimates, one that subtracts the velocity gradient and a second that does not subtract the velocity gradient. We adopt the momentum estimate by model MF, which subtracts the velocity gradients as the fiducial estimate. The trend between the momentum driven by outflows and the number of YSOs is similar to that of the mass estimates for four clouds. In Ophiuchus, the 1D momentum driven by outflows is around 10\% of the total 1D momentum of the host cloud. In Taurus, this momentum ratio is 8\%. In Perseus and Orion, the 1D momentum driven by outflows is around 16\% of the total 1D momentum, which indicates there are likely more high velocity structures identified as outflows. This is consistent with the star formation history of these clouds. Taurus has a relatively low star formation rate and mainly forms low mass stars. 
On the other hand, Ophiuchus and Orion are home to clusters forming higher-mass stars and have more active star formation. Perseus also contains some intermediate mass stars, including a couple of B-type stars \citep{2011ApJ...742..105A}. 

Ophiuchus, Perseus and Orion have larger velocity gradients than Taurus, which indicates these clouds contain more complex cloud kinematic structure. These velocity gradients are likely caused by large scale processes, such as gas accretion \citep{2010A&A...520A..17K} or converging atomic gas flows caused by spiral density waves \citep{1995ApJ...443..152W,2006ApJ...648.1052H}. Given the magnitude of these bulk motions, we expect that the momentum estimate that includes the velocity gradient overestimates the contribution of outflows, while the momentum that subtracts the velocity gradient provides a better estimate.

\begin{figure*}[hbt!]
\centering
\includegraphics[width=0.98\linewidth]{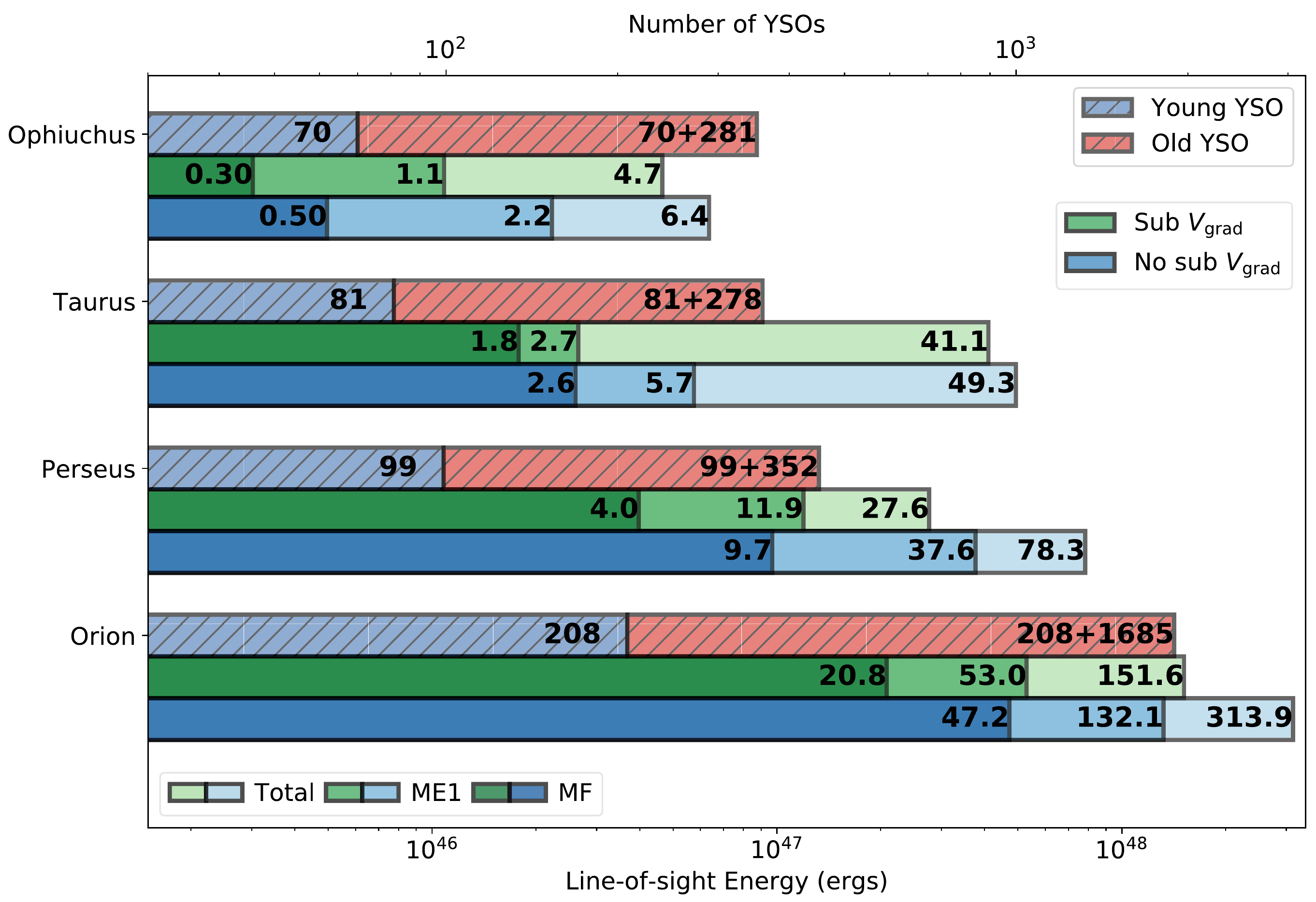}
\caption{Outflow line-of-sight kinetic energy (solid colors) and the number of YSOs (hatched colors) in the four molecular clouds. Numbers in the hatched bars indicate the number of YSOs. Numbers in the unhatched bars indicate the kinetic energy in units of $10^{46}$ ergs. The total energy indicates the energy of the host cloud within the area studied.}
\label{fig.stat-energy-all}
\end{figure*}

Figure~\ref{fig.stat-energy-all} shows the LOS outflow kinetic energy for the four molecular clouds. We adopt the kinetic energy estimates by model MF that subtracts the velocity gradient for the comparisons. In Ophiuchus, the LOS kinetic energy injected by outflows is around 6\% of the total LOS kinetic energy of the host cloud. In Taurus, this energy ratio is 4\%. In Perseus and Orion, the LOS kinetic energy injected by outflows is around 14\% of the total LOS kinetic energy. Since kinetic energy is proportional to $v^{2}$, one caveat here is that a small amount of mass located at high velocities likely dominates the kinetic energy estimates. However, due to observational limits, some high velocity gas emission vanishes into background noise, such that our kinetic energy calculation is likely an underestimate. As discussed in \citet{2020ApJ...905..172X}, this could potentially introduce a factor of two uncertainty.


\subsection{Correlation Between the Outflow Properties and the Number of YSOs}
\label{Correlation Between the Outflow Properties and the Number of YSOs}

In this section, we study the correlation between the total mass, momentum and energy associated with outflows and the number of YSOs. Due to the extended and overlapping nature of the outflow emission it is not possible to definitively associate particular sources with particular outflows. Instead, we consider an approach that independently counts the number of YSOs and the outflow impact in a region. We define a ``window'' to scan through the entire cloud region. We adopt window sizes of 0.5~pc$\times$0.5~pc, 1~pc$\times$1~pc and 2~pc$\times$2~pc to scan Ophiuchus, and 2~pc$\times$2~pc, 3~pc$\times$3~pc and 5~pc$\times$5~pc to scan Taurus and Orion. Since Perseus has the poorest physical resolution, and its full map is narrow, we only adopt window sizes of 2~pc$\times$2~pc and 3~pc$\times$3~pc to prevent the window exceeding the region. We examine the effect of our choice of window sizes in Appendix~\ref{Discussion on the Effect of Box Sizes on the YSO-Mass Relation}. We set a scanning step size such that each box overlaps with its neighbors by at least 80\%. We calculate the mass, momentum and energy inside the window and count the number of YSOs inside. We consider the outflow relationship between the number of both young YSOs and total YSOs.

\begin{figure*}[hbt!]
\centering
\includegraphics[width=0.98\linewidth]{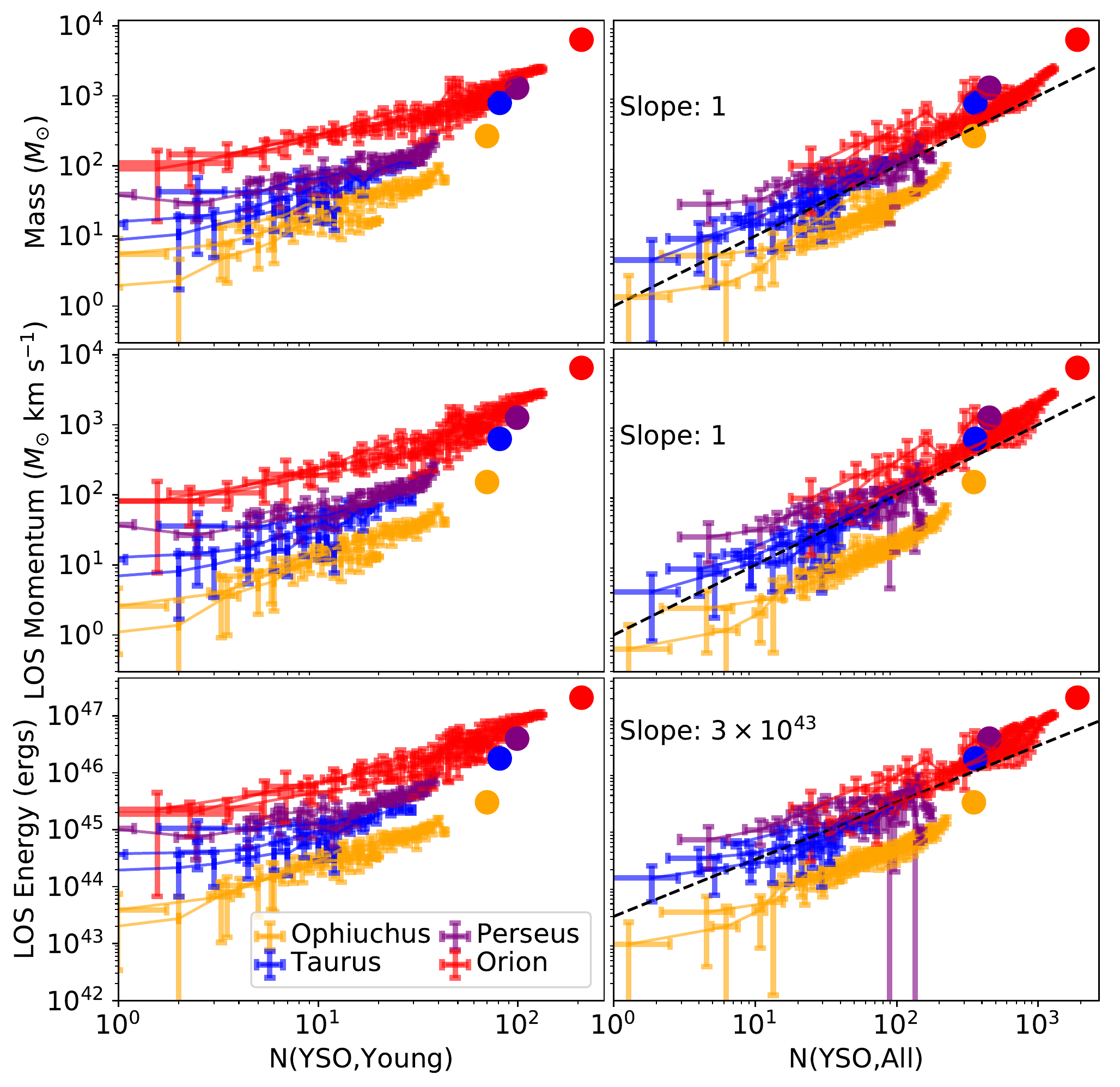}

\caption{Correlation between the mass (top row)/momentum (middle row)/energy (bottom row) associated with feedback and the number of young YSOs (left panel) or the number of all YSOs (right panel) in the four clouds. The filled circles indicate the total outflow mass (top row)/momentum (middle row)/energy (bottom row) and the total number of YSOs (young YSOs in the left panel, all YSOs in the right panel) in the four clouds. The window sizes are 0.5~pc$\times$0.5~pc, 1~pc$\times$1~pc and 2~pc$\times$2~pc for Ophiuchus, and 2~pc$\times$2~pc, 3~pc$\times$3~pc and 5~pc$\times$5~pc for Taurus and Orion. The window sizes for Perseus are 2~pc$\times$2~pc and 3~pc$\times$3~pc.
}
\label{fig.stat-yso-mass-yso}
\end{figure*}

Figure~\ref{fig.stat-yso-mass-yso} shows the correlation between the mass associated with feedback and the number of YSOs in the four clouds. The outflow mass grows linearly with the number of young YSOs for all four clouds. However, the clouds have different offsets, indicating that the outflow mass per young YSO varies. Orion has the most mass associated with outflows per young YSO. However, Orion also has a very large population of older YSOs, which may also drive outflows. The right panel of Figure~\ref{fig.stat-yso-mass-yso} shows that the correlation between the outflow mass and the number of total YSOs is tighter, and most of the separation between different clouds disappears. There is a linear relation between the outflow mass and the number of all YSOs, where on average, each YSO is associated with one \msun\ of outflow material. The outflow mass directly launched by the protostar is expected to be 10-30\% of the accreted gas, i.e., 10-30\% of the star mass \citep[e.g.,][]{1988ApJ...328L..19S,1992ApJ...394..117P}. However, here our outflow estimate also counts entrained mass, which is much higher \citep[e.g.,][]{2017ApJ...847..104O}. Our results suggest one solar mass of outflow material per source. Assuming a one solar mass star for example, theoretical models for outflow launching predict that the outflow mass directly launched would be about 0.1-0.3 \msun \citep[e.g.,][]{1994ApJ...429..781S,1996A&A...311..858B}. This implies that 0.7-0.9 \msun\ of the outflow gas we identify is entrained. This gives a mass-loading factor of 2.3-9. This is comparable to the mass-loading factor estimated from simulations of individual protostars and the average core-mass-function to stellar IMF offset \citep{2013MNRAS.431.1719M,2014ApJ...784...61O,2017ApJ...847..104O}. {It is worth noting that our outflow estimate is larger than previous observational estimates. The main reason is that \CASItD\ is able to capture the outflow material that has relatively low velocities. The low-velocity outflowing gas entrains a significant amount of mass that also contributes to the momentum and kinetic energy of the host cloud, but it is not possible to separate this gas from the ambient gas by eye. Previous work  mainly considered the high-velocity outflow mass only. }


Younger sources power stronger outflows \citep{2016ARA&A..54..491B}, so we might expect a tighter relation between the outflow mass and the number of young YSOs. However, we see better correlation between outflow mass and the total YSO number. There are several possible explanations. First, low-mass sources, with age up to 3 to 4 Myr \citep{2016ARA&A..54..491B} continue to launch outflows, albeit weaker ones. We discuss an example in Section~\ref{Outflows without Driving Sources and False Detections}, where an outflow is likely driven by old evolutionary stage YSOs. Second, although young YSOs have a higher mass-loss rate, the outflow mass is a combination of both mass-loss rate and the launching duration. The older YSOs likely have ejected a significant amount of gas by the time they are observed. In addition, our method may include outflow ``relics.'' The dissipation timescale of outflow features produced by young stars is much longer than the lifetime of the driving source. These fossil outflows might remain even as their driving sources evolve to older evolutionary stages. \citet{2006ApJ...653..416C} found a similar scenario in numerical simulations, where fossil outflows retain speeds above the turbulent velocity for a timescale that is 10 times the duration of their driving source. Similarly, \citet{2017ApJ...847..104O} found in simulations that although the high accretion and active outflow phase lasts for $\sim$ 0.05 Myr, their impact on the velocity dispersion of the host clouds remains for several 0.1 Myr. 

{In our outflow mass estimate, a significant amount of the total outflow mass derives from emission in the cloud velocity channels ($|v| \lesssim 1$ \kms). This raises the concern that our results may be contaminated by dense, non-outflow material. In Appendix~\ref{Correlation Between the Gas Mass and the Number of YSOs} we evaluate the effect of dense gas contamination on the derived outflow masses and YSO relations. We show that our outflow mass estimate is not significantly contaminated by dense gas. 

}

We show the correlation between the outflow momentum and the number of YSOs in Figure~\ref{fig.stat-yso-mass-yso}. The trend is similar to that of outflow mass. On average, one YSO injects 1 $\rm{M_{\sun} km/s}$. This indicates the mass weighted LOS velocity is 1~km/s, which is even smaller than the cloud turbulent velocity derived from \co. This is because the \13co\ emitting region dominates the mass of the cloud and the outflows, which usually has a velocity dispersion only half that of \co. For example, after subtracting the velocity gradients, the \co\ velocity dispersion in Orion is $\sim$1.5~km/s, while that of \13co\ is only 0.8 km/s. The mass-weighted LOS velocity dispersion in Orion is 0.98 km/s. Model MF predicted outflow gas has a LOS velocity dispersion of 1.02 km/s in Orion. All these numbers are similar for the four clouds. {Although the outflow \co\ emission spans 2-3 km/s across the velocity channels, most of the mass is associated with gas within 0.8 km/s of the cloud mean velocity, i.e., the \13co\ emitting regions.} {Near the rest-frame velocity channels, the gas mass increases significantly but the fraction of gas associated with feedback drops. Consequently, there is a competition between these two effects. \citet{2020ApJ...905..172X} 
examined the outflow mass located near the rest-frame velocity, which was ignored in previous outflow surveys.
As shown in Figure 16 and 17 in \citet{2020ApJ...905..172X}, the rest-frame gas that is associated with the outflows accounts for almost 75\% of the total outflow mass. Thus, the mass-weighted LOS velocity is relatively low, because high-velocity gas contributes only a small portion of the total outflow mass. 
} In addition, the resolution and low-signal-to-noise of the data makes detecting high-velocity outflow emission difficult.  Consequently, the mass weighted LOS velocity of outflows is only 1 km/s for all four clouds. We show a position-velocity diagram of an outflow in \13co\ in Appendix~\ref{Outflow Morphology}.

Our mass-weighted LOS outflow velocity of 1 km/s is low
compared to typical outflow velocities estimated in previous works of 1-4 km/s \citep[e.g.][]{2010ApJ...715.1170A}. This is caused by the mass-weighting. In most cases, observers discard the rest-frame gas emission, whose outflow morphology is difficult to identify. This leads to a higher LOS velocity estimate. However, \CASItD\ is able to pick out the outflow emission from the confused ambient gas emission near the rest-frame velocity. Although the fraction of mass associated with feedback is low, perhaps a few percent, the total mass near rest-frame, especially in \13co\ emission channels, is significantly higher than that located at high-velocity channels. After correcting for the contamination, the outflow gas near the rest frame still dominates the total mass, which is consistent with the conclusion in \citet{2020ApJ...905..172X}. On the other hand, high-velocity gas is less dense and may be lost in the background noise. Thus, we likely underestimate the total outflow momentum by at least 10\% \citet{2020ApJ...905..172X}.

The correlation between the outflow kinetic energy and the number of YSOs in the four clouds is similar to that of mass, as shown in Figure~\ref{fig.stat-yso-mass-yso}.


\subsection{Quantifying the Impact of Feedback with Turbulent Statistics}
\label{Quantifying the Impact of Feedback with Turbulent Statistics}

Feedback, including stellar winds and outflows, injects kinetic energy into the host cloud. The input energy influences the shape of spatial power spectrum (SPS) of the integrated intensity map of \co. \citet{2020ApJ...890...64X} found that stellar wind-generated bubbles flatten the SPS of the \13co\ integrated intensity map, which indicates mass and energy are injected at small scales. In this section, we investigate how protostellar outflows affect the SPS of the \co\ integrated intensity maps. 

The SPS is defined as the square of the 2D Fourier transform of an image:
\begin{equation}
\label{sps-eq1}
\begin{split}
\mathcal{P}(k) & =\sum_{|\vec{k}|=k}|\mathcal{M}_{0}({ \vec{k}})|^{2} \\
 & =|\int_{-\infty}^{\infty}\int_{-\infty}^{\infty}M_{0}({\vec{x}})e^{-2\pi j{\vec{k}\vec{x}}}d{\vec{x}}|^2,
\end{split}
\end{equation}
where $\mathcal{M}_{0}$ is the 0$^{th}$ moment (integrated intensity) of \co.

\begin{figure*}[hbt!]
\centering
\includegraphics[width=0.97\linewidth]{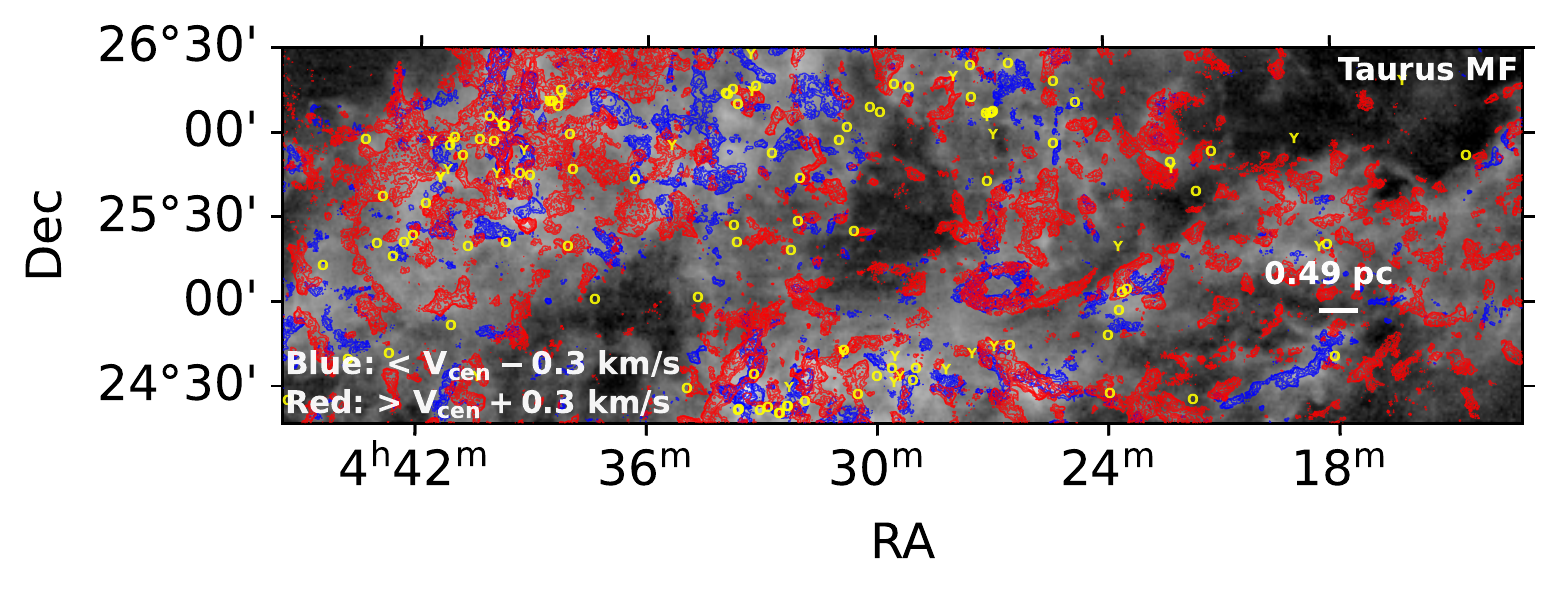}
\includegraphics[width=0.97\linewidth]{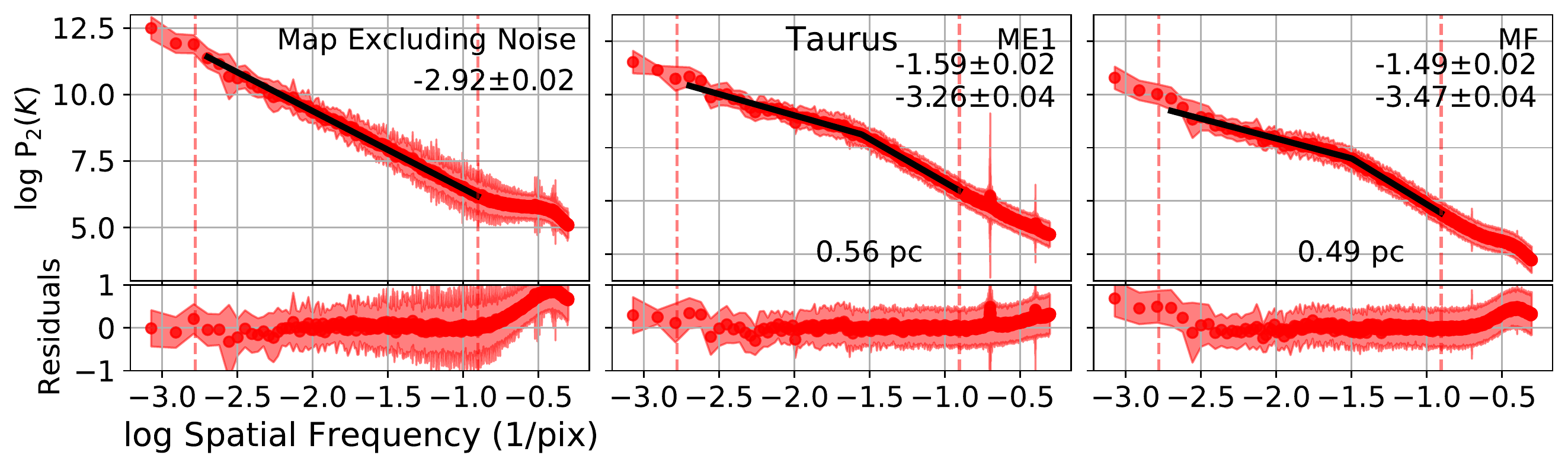}

\caption{Upper panel: intensity of \co\ (1-0) integrated over all velocity channels for a subregion in Taurus, overlaid with the model MF prediction (red and blue contours). The physical scale of the break point in the SPS is also shown for reference. Lower panels: the spatial power spectrum (SPS) of the regions in Taurus where the emission is above 0.2 K (excluding noise) and the SPS of the model ME1 and MF predicted feedback regions. The top right number in each panel indicates the slope of the fit. The middle bottom number in each panel indicates the physical scale of the broken point.  }
\label{fig.ps-moment-0-taurus}
\end{figure*}

Figure~\ref{fig.ps-moment-0-taurus} shows the SPS of Taurus where the emission is above 0.2 K (i.e., excluding noise) and the SPS of the model ME1 and MF predicted feedback regions. Figure~\ref{fig.ps-moment-0-taurus} shows that the SPS of the feedback predictions breaks into two power laws. The break point corresponds to a physical scale of $\sim 0.5$~pc, which might indicate the typical outflow mass and energy injection scale. For reference, we show the scale bar of this injection length in a subregion of Taurus in Figure~\ref{fig.ps-moment-0-taurus}. This outflow injection scale is comparable to the typical outflow size in Taurus. 


We find similar broken power laws in Ophiuchus, Perseus and Orion. We list the SPS fits for all four regions in Table~\ref{tab-powerspec-sum}. 
For a given cloud, we find the model ME1 break point scale to be similar and usually  slightly higher, than that of model MF. Table~{1} 
also lists the median of the distance between young YSOs and their four nearest young companions.

There are several competing effects that likely influence the location of the break point. For isolated outflows, the break point size is close to the outflow physical scale. However, the typical outflow size is not necessarily the same in all regions. It is influenced by the gas density and how readily the outflow can expand, as well as the source mass and age. In very clustered regions, outflows interact with each other, since the separation between sources is smaller. This causes 
the outflow emission to blend together, which increases the break point. 

The physical scale of the break point is small for Orion and Ophiuchus, and Taurus has a intermediate break point scale, while Perseus has the largest. Among the four clouds, Taurus has the most distributed and least clustered YSOs. Taurus also has the lowest average gas density. The mass/energy injection scale in Taurus reflects the physical size of individual molecular outflows. While in Orion and Ophiuchus, the average distance between stars is small, which increases the probability that outflows from different sources interact. The mean gas densities are also higher, and 
high external pressure from the surrounding gas may also act to limit the propagation of outflows. For example, \citet{2017ApJ...846..144K} found all dense cores in Orion are pressure confined. All of these effects help to explain the relatively small mass/energy injection scale. 

The physical scale of the break point in Perseus is significantly larger than that of the other regions. This might be due to both its typical density, which is lower than that of Orion and Ophiuchus, and the presence of multiple clusters in Perseus. Perseus hosts several intermediate mass star-forming clusters, such as NGC 1333 and IC348. 
Outflows in these clusters interact and blend together, such that mass/energy injection occurs on a relatively larger scale compared to individual stars. 
The average young YSO separation is slightly larger in Perseus than that in Orion and Ophiuchus, in part because the two main clusters contain few young YSOs.
Consequently, the combination of these effects may explain the large mass/energy injection scale in Perseus.

To investigate how YSO clusters affect the SPS break point, we conduct a SPS analysis towards two star clusters, NGC1333 in Perseus and L1688 in Ophiuchus. We adopt different window sizes to explore the location of the break point in the cluster when viewed on different size scales. When the window size is smaller than the cluster size, we find there is no break point in the SPS for both clustered region and the slope is steeper. When the window size is two or more times larger than the cluster size, the break point appears. However, the break point scale is not neatly correlated with the size of the cluster nor the window size. NGC1333 has a break point between 0.6 pc and 0.7 pc, while L1688 has a break point scale between 0.3 pc and 0.4 pc. These scales are similar to those of the full predictions, which suggests that the outflows in the clustered regions are influencing the location of the overall break point.

We also explore the effect of different 
observation resolutions on the results. The observations of Ophiuchus, Taurus and Orion have a similar physical scale per pixel, $0.013-0.015$~pc/pixel. However, the physical scale per pixel of Perseus is twice that of the others. We convolve the Ophiuchus and Orion observations to a similar effective resolution as Perseus, and apply the \CASItD\ models to these convolved data cubes and conduct the same SPS analysis on the convolved data. However, we find only minor changes in the break point. For example, the break point remains the same in the convolved Ophiuchus model MF prediction and increases by 0.1 pc in the convolved Ophiuchus model ME1 prediction. However, it decreases by 0.08 pc in the convolved Orion ME1 prediction, while increasing by 0.08 pc  in the Orion MF prediction. We find no significant trend when convolving with a larger beam and conclude the analysis is relatively insensitive to the resolution. However, this test suggests that the break point has an uncertainty of $\pm0.1$ pc, which is smaller than the difference between the break point of Perseus and those of the other clouds. 

{Finally, to test whether (non-feedback) gas structures influence the break point location, we 
compute the SPS for the dense regions of the clouds that are traced by \13co. We mask out the low intensity pixels to isolate only the dense gas. We compare the results for two clouds, Orion and Taurus. When all pixels with intensities below 5 K are removed, there is no break point in the Orion SPS. For the densest regions, i.e., as defined using an intensity cutoff of 10 K, we find a break point at 0.6 pc. This is two times larger than that of the break point of the outflow maps. We find a similar result in Taurus. When applying a high intensity cut-off, we find a break point around 1 pc, which is a size-scale twice that of the break point identified by the Taurus SPS outflow analysis.
We conclude that our outflow predictions are not correlated with dense cloud structures. 
}

\begin{table*}[]
\begin{center}
  \label{tab-powerspec-sum}
  \caption{Fitting Results of the Spatial Power Spectrum}
 \begin{tabular}{|c|c|c|c|c|c||c|}
\cline{1-7}
    \multirow{2}[0]{*}{Cloud} & \multicolumn{1}{c|}{Full Map$^{a}$} & \multicolumn{2}{c|}{ME1$^{b}$} & \multicolumn{2}{c||}{MF$^{b}$} & \multicolumn{1}{c|}{Young YSO}  \\ \cline{2-6}

          & \multicolumn{1}{c|}{Slope} & \multicolumn{1}{c|}{Slope} & \multicolumn{1}{c|}{Break (pc)} & \multicolumn{1}{c|}{Slope} & \multicolumn{1}{c||}{Break (pc)} &\multicolumn{1}{c|}{Seperation$^{c}$ (pc)} \\ \cline{1-7}

    \multirow{2}[0]{*}{Ophiuchus} & \multirow{2}[0]{*}{-3.17$\pm$0.02} & -2.47$\pm$0.05 & \multirow{2}[0]{*}{0.36} & -1.82$\pm$0.05 & \multirow{2}[0]{*}{0.38} &  \multirow{2}[0]{*}{0.24} \\

          &       & -3.71$\pm$0.07 &       & -3.78$\pm$0.08 &   &   \\ \cline{1-7}

    \multirow{2}[0]{*}{Taurus} & \multirow{2}[0]{*}{-2.92$\pm$0.02} & -1.59$\pm$0.02 & \multirow{2}[0]{*}{0.56} & -1.49$\pm$0.02 & \multirow{2}[0]{*}{0.49}  &  \multirow{2}[0]{*}{0.85} \\

          &       & -3.26$\pm$0.04 &       & -3.47$\pm$0.04 &   &   \\ \cline{1-7}

    \multirow{2}[0]{*}{Perseus} & \multirow{2}[0]{*}{-3.07$\pm$0.02} & -2.18$\pm$0.04 & \multirow{2}[0]{*}{1.03} & -2.21$\pm$0.03 & \multirow{2}[0]{*}{0.65}  &  \multirow{2}[0]{*}{0.31} \\

          &       & -3.40$\pm$0.06  &       & -3.47$\pm$0.05 &   &   \\ \cline{1-7}

    \multirow{2}[0]{*}{Orion} & \multirow{2}[0]{*}{-2.95$\pm$0.02} & -2.29$\pm$0.03 & \multirow{2}[0]{*}{0.36} & -2.06$\pm$0.03 & \multirow{2}[0]{*}{0.27}  &   \multirow{2}[0]{*}{0.22} \\

          &       & -3.75$\pm$0.05 &       & -4.83$\pm$0.05 &   &   \\ 
\cline{1-7} 

\multicolumn{7}{p{0.87\linewidth}}{Notes:}\\
\multicolumn{7}{p{0.87\linewidth}}{
$^{a}$ Single power-law fit results for the spatial power spectra for the emission regions (excluding noise).}\\
\multicolumn{7}{p{0.87\linewidth}}{
$^{b}$ Broken power-law fit results for the spatial power spectra applied to the ME1 and MF feedback maps.
}\\
\multicolumn{7}{p{0.87\linewidth}}{
$^{c}$ Median of the separation between YSOs and their four nearest companions.
}
\end{tabular}%
\end{center}
\end{table*}%

\section{Discussion}
\label{Discussion}

\subsection{Outflows without Driving Sources and False Detections}
\label{Outflows without Driving Sources and False Detections}

Confirming outflow identifications requires ancillary data, e.g., YSO catalogs. With the help of YSO locations, we are able to increase confidence in the \CASItD\ predictions, which are based on \co\ data only. In this section, we discuss two cases predicted by \CASItD: a region in Taurus with no detected YSOs, and a region in Orion with no young YSOs.

\begin{figure*}[hbt!]
\centering
\includegraphics[width=0.98\linewidth]{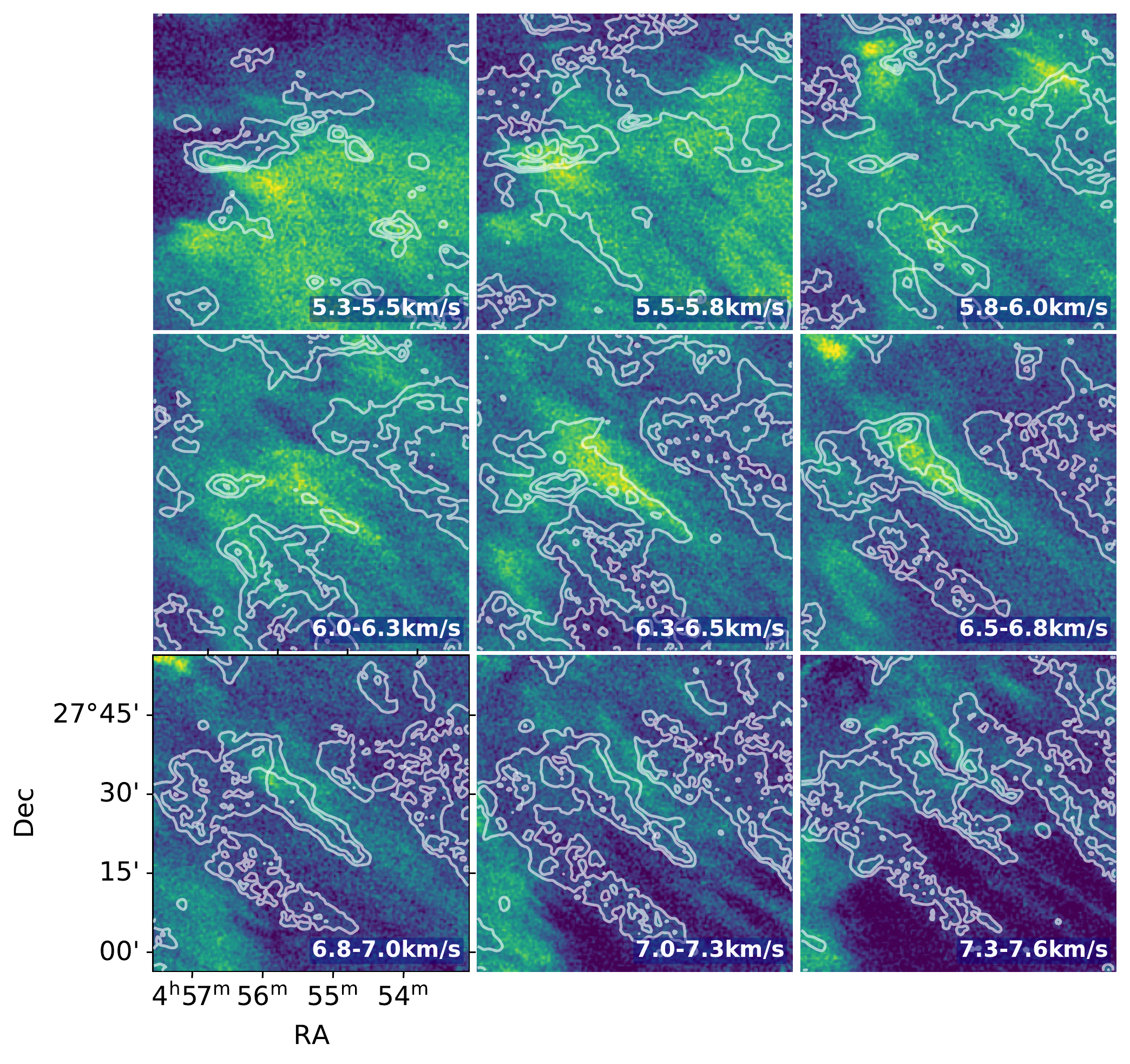}
\caption{\co\ channel map of a region with no YSOs in Taurus overlaid with the prediction by model MF in white contours. }
\label{fig.channel-map-taurus-mf}
\end{figure*}

Figure~\ref{fig.channel-map-taurus-mf} presents the channel by channel prediction by model MF on a region with no YSOs in Taurus. This region is considered to be a trans-Alfvenic flow regulated by magnetic fields \citep{2012MNRAS.420.1562H,2016MNRAS.461.3918H}, which has low surface brightness \co\ emission. There are no known YSOs observed by GAIA or Herschel \citep{2020A&A...638A..85R}. Consequently, we believe there is no outflow activity. However, model MF predicts outflow activity in this region. Inspection of the channel map shows that a coherent jet-like structure exists across several velocity channels. These structures actually visually resemble outflows, which might be reason for the model failure. 

\begin{figure*}[hbt!]
\centering
\includegraphics[width=0.98\linewidth]{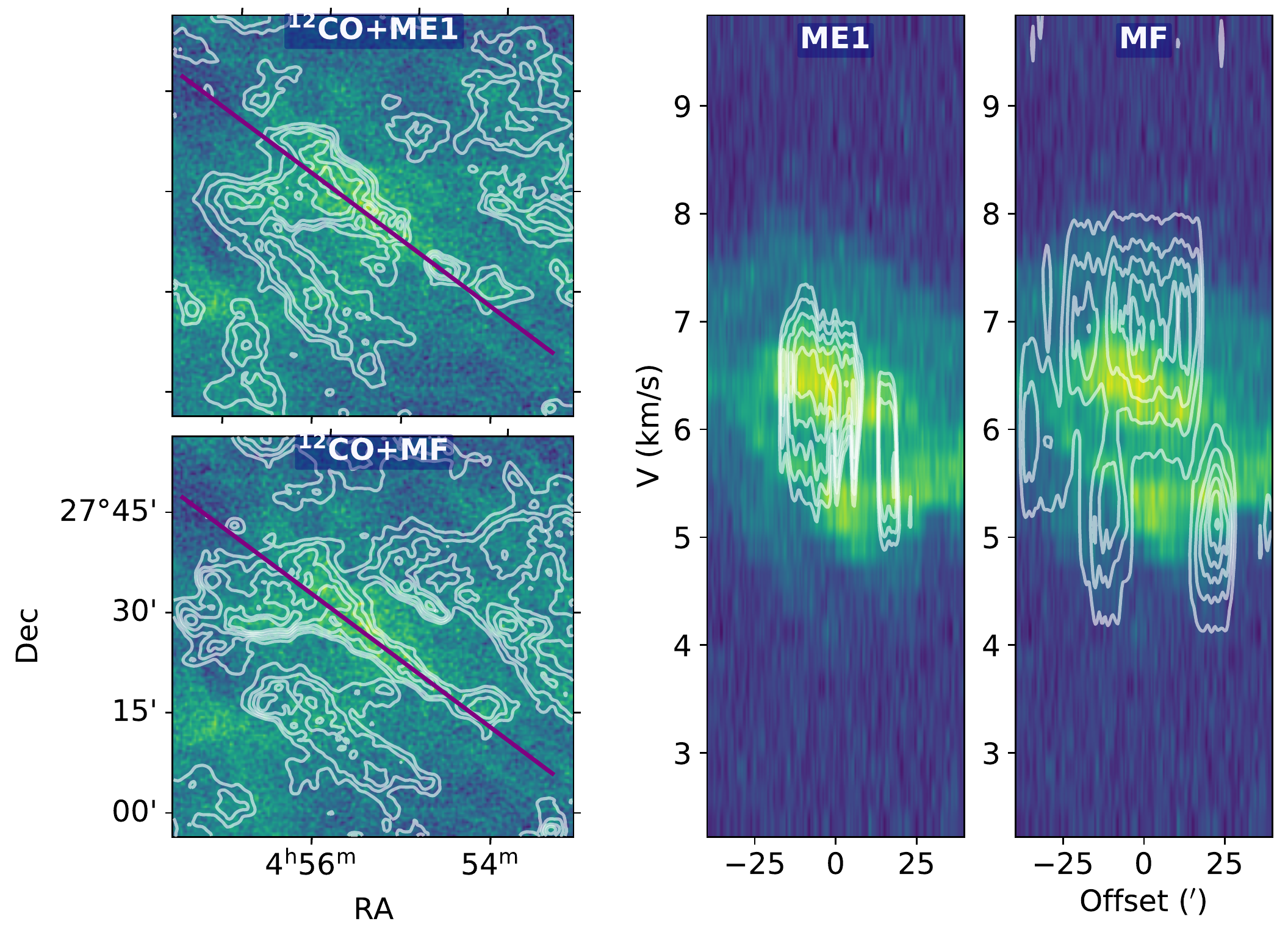}
\caption{Position-velocity diagram of \co\ emission toward a region with no YSOs in Taurus. Left panel: integrated intensity of \co\ over the the full velocity range (from -1.5 km/s to 13.4 km/s) overlaid with the model ME1 and MF predictions in white contours. The purple line illustrates the cut direction of the position-velocity diagram. Middle and right panel: position-velocity diagram of \co\ emission overlaid with the model ME1 and MF predictions in white contours.  }
\label{fig.pv-map-taurus-failcase}
\end{figure*}

Figure~\ref{fig.pv-map-taurus-failcase} shows the model MF prediction along a position-velocity cut of the fake outflow in Figure~\ref{fig.channel-map-taurus-mf}. In the position-velocity diagram, we identify two faint coherent high velocity blobs between 4 and 5 km/s, Their morphology is similar to some of the faint outflows \citep[e.g., TMO22 in][]{2015ApJS..219...20L}. Since these \co\ structures are indistinguishable from true outflows, \CASItD\ is not able to recognize these false detections based only on \co\ morphology. In terms of morphology, \CASItD\ performs robustly in identifying coherent high velocity features that are similar to outflows. But many mechanisms may cause coherent high velocity features, including but not limited to cloud formation, cloud-cloud collision, gas phase transition near the cloud boundary or gas flows regulated by MHD waves \citep{2012MNRAS.420.1562H,2014A&A...571A..32M,2014ApJ...791L..23N,2016MNRAS.461.3918H}.

\begin{figure*}[hbt!]
\centering
\includegraphics[width=0.98\linewidth]{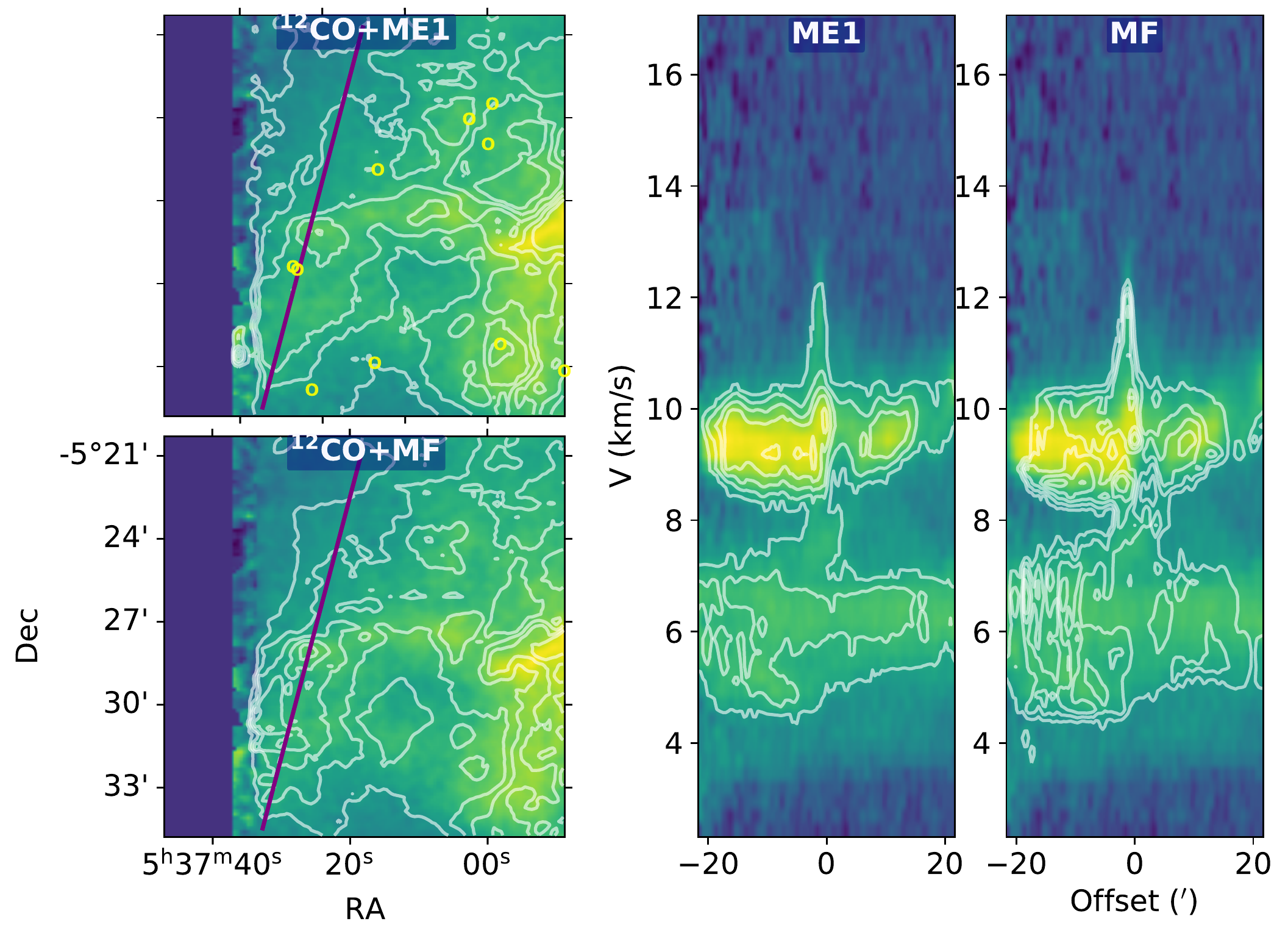}
\caption{Position-velocity diagram of \co\ emission toward a region without young YSOs in Orion. Left panel: integrated intensity of \co\ over the full velocity range (from 2.3 km/s to 17.1 km/s) overlaid with the model ME1 and MF predictions in white contours. Letters ``O'' mark old YSO positions. The purple line illustrates the cut direction of the position-velocity diagram. Middle and right panel: position-velocity diagram of \co\ emission overlaid with the model ME1 and MF predictions in white contours. }
\label{fig.pv-map-orion-case2}
\end{figure*}

Figure~\ref{fig.pv-map-orion-case2} shows an example of the performance of models ME1 and MF toward a region where there are no young YSOs in Orion. In the position-velocity diagram, we identify a clear outflow-like feature above 10~km/s, which is also identified by the two models. We notice that there are several old evolutionary stage YSOs nearby, which might act as driving sources. One reason might be that the boundary between young and old is not so well defined, which is discussed in Section ~\ref{Correlation Between the Outflow Properties and the Number of YSOs}. Or these outflow structures might exist for a much longer timescale than that of their driving sources. We recognize that there are two velocity components in the molecular cloud at this position. A narrow gas bridge located around 8~km/s connects these two gas components, which is likely an outflow. We cannot confirm the origin of this high velocity component. It might be caused by the fossil outflows from YSOs or by two converging gas flows. All these mechanisms cause similar high-velocity features, which cannot be easily distinguished either visually or using \CASItD.

{In order to validate that our \CASItD\ models are sensitive to gas velocity structures and morphology but not only the dense regions, we examine the \CASItD\ performance on several previously identified filaments in Taurus. \citet{2014MNRAS.444.2507P} identified 10 filametary structures in \13co\ emission in Taurus, and illustrated the PPV diagrams of the filaments in Figure~15 in their paper. The filaments have very coherent motions and exhibit small velocity dispersions. Their filaments 3, 4, 5, 6, and 7 do not have significant high-velocity components. When comparing with our outflow predictions (Figure~\ref{fig.pred-taurus-lm-ME1} and \ref{fig.pred-taurus-lm-MF} in this work) and bubble predictions \citep[Figure 17 and 18 in][]{2020ApJ...890...64X}, we find that both models predict little feedback along these filaments. However, when we look at Filament 2 (also known as L1495/B213), we see a clear high-velocity structure with a ``U" shape. This morphology is consistent with the theoretical bubble morphology shown in Figure~5 in \citet{2011ApJ...742..105A}. Unsurprisingly, both \CASItD\ models that identify bubbles and outflows predict the presence of feedback at the location of filament 2.  \CASItD\ may return a false detection if the morphology of a structure is similar to that of outflows or bubbles. In this case, visual inspection suggests that filament 2 also contains a bubble, so it is likely that the feedback identified by our models at this location is real.} We emphasize that machine learning models are not perfect tools. They must be applied with caution and checked thoroughly.

\subsection{Outflows versus Bubbles}
\label{Outflows v.s Bubbles}

In this section we investigate the overlap between a model identifying outflow feedback and one identifying stellar wind feedback. We present the prediction by two sets of models: those trained to identify protostellar outflows and those trained to identify stellar wind driven bubbles. 

\begin{figure}[hbt!]
\centering
\includegraphics[width=0.98\linewidth]{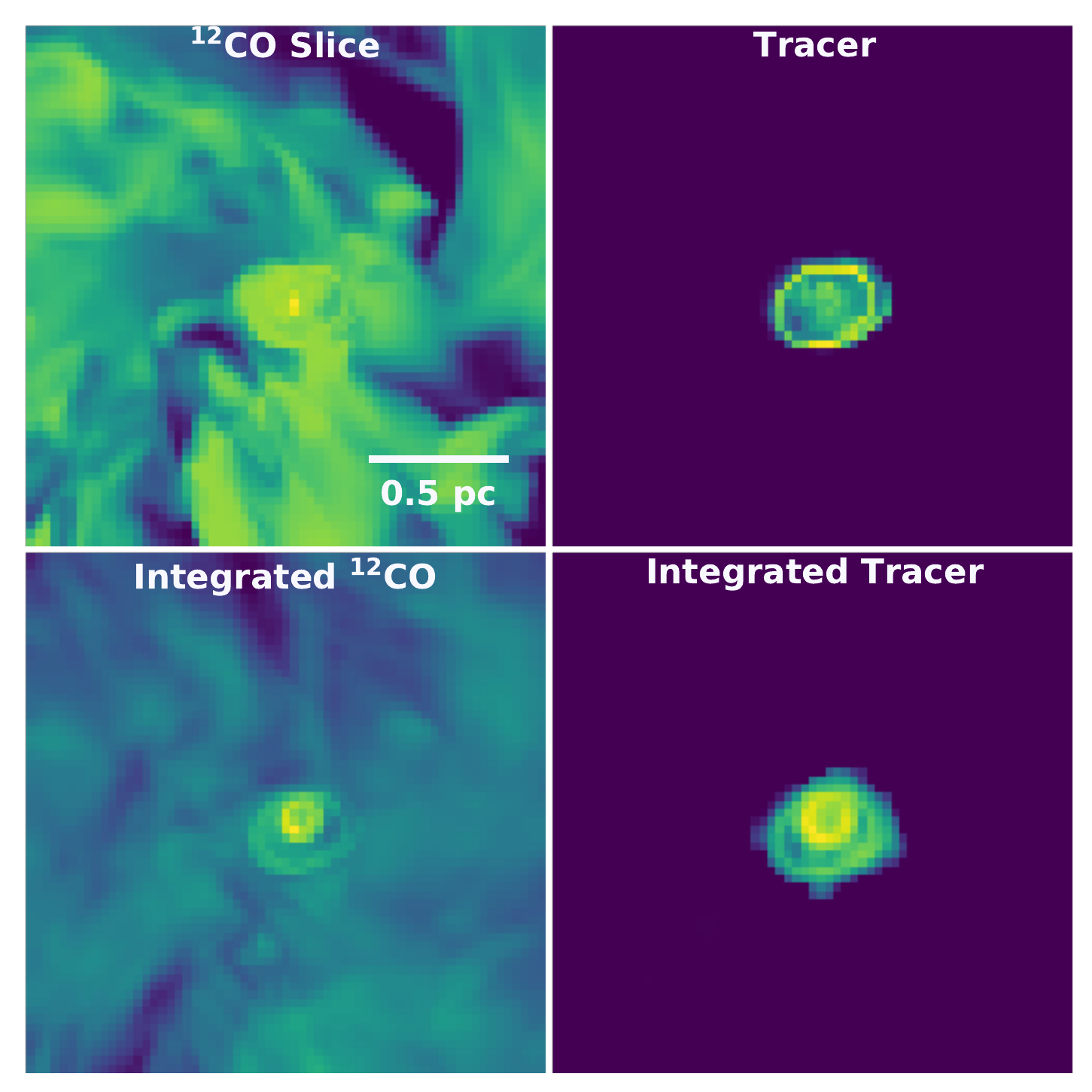}
\caption{A synthetic outflow whose launching axis is parallel to the line of sight. }
\label{fig.outflow-los-synth}
\end{figure}

The morphology of bubbles is more symmetric and arc-like compared to that of outflows. However, when the line of sight is parallel to an outflow launching axis, i.e., looking through the cavity of the outflow, the morphology of the outflow can resemble that of a bubble. Figure~\ref{fig.outflow-los-synth} illustrates an example of a synthetic outflow whose launching axis is parallel to the line of sight. Consequently, models trained to identify outflows are likely to also identify some bubble structures. 

\begin{figure*}[hbt!]
\centering
\includegraphics[width=0.98\linewidth]{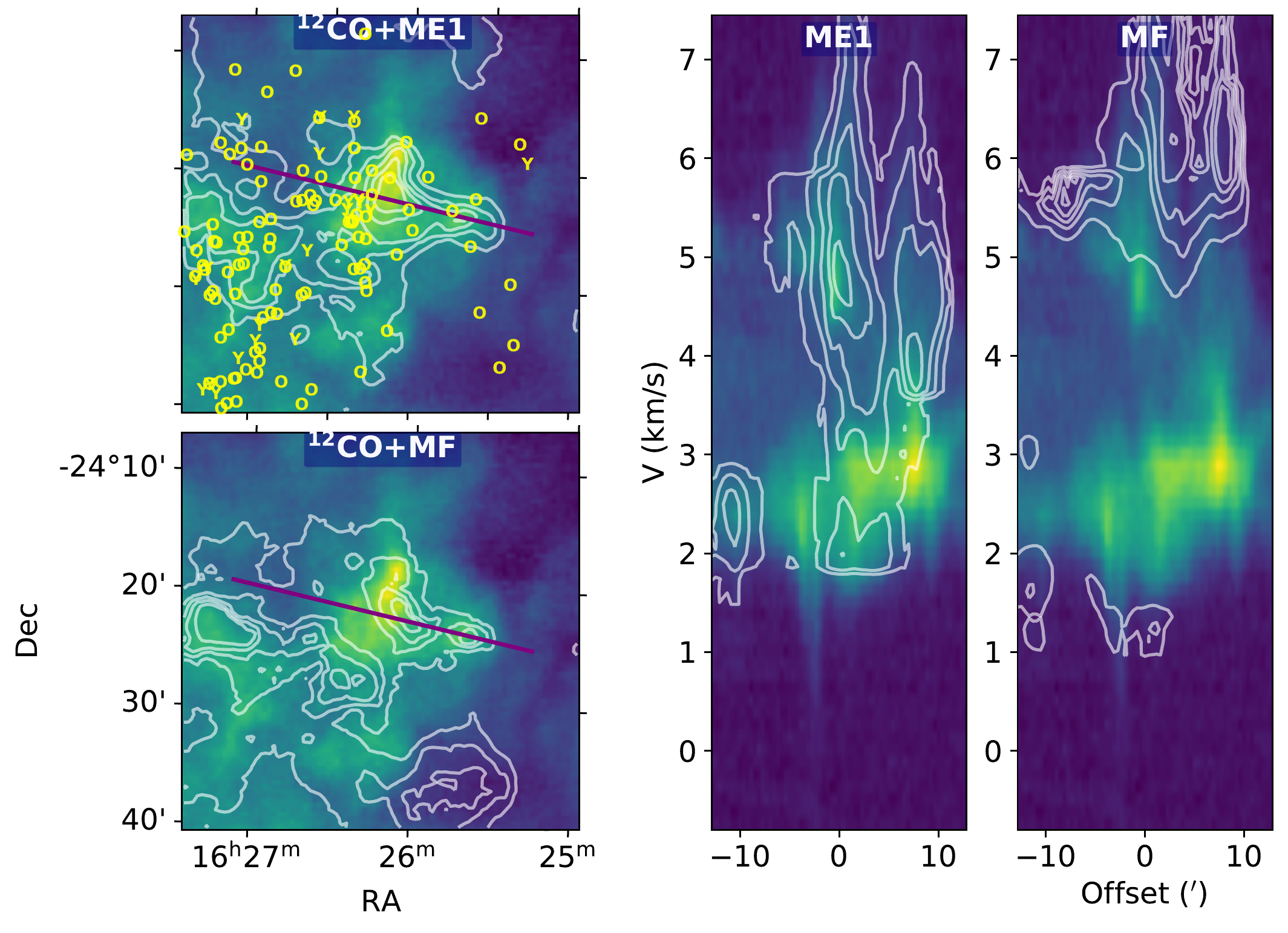}
\caption{Position-velocity diagram of \co\ emission towards Ophiuchus L1688 Region, similar to Figure~\ref{fig.pv-map-oph-goodcase} but predicted by models trained to identify stellar wind driven bubbles. }
\label{fig.pv-map-oph-bubble}
\end{figure*}

Figure~\ref{fig.pv-map-oph-bubble} shows the prediction by models trained to identify stellar wind driven bubbles towards Ophiuchus L1688 Region. Comparison with Figure~\ref{fig.pv-map-oph-goodcase} shows that the prediction by models trained to identify protostellar outflows covers most of the regions that are identified as bubbles. Quantitively, 68\% of the voxels are identified as feedback by both MF models, 21\% of the voxels are identified as only outflow feedback, while 11\% of the voxels are identified as only belonging to bubbles. Statistically, we might neglect 13\% of the feedback gas associated with stellar winds when adopting the mass estimates predicted by models trained to identify protostellar outflows. On average, the outflow mass predicted by models can serve as a decent estimate for feedback mass.

To better investigate the difference between outflows and bubbles, we employ an unsupervised learning method t-SNE (t-distributed Stochastic Neighbor Embedding) to study the similarities between the spectra of outflows and that of bubbles. We discuss t-SNE and the results in detail in Appendix~\ref{Spectra of Outflows and Bubbles}. Our main finding is that there is a variation in spectra shape for outflows and bubbles. And even the spectra that are dominated by outflow gas show various spectral shapes. Meanwhile, there are some ``transitional spectra'' with a shape similar to both. We note that the t-SNE representation encodes only 1D information about the spectra and does not represent any correlations in the spatial dimension. Consequently, this analysis provides intuition for understanding the features that the models select when they identify feedback, and it illustrates the challenge of distinguishing outflows and bubbles based only on the LOS spectrum.

\subsection{Impact of Outflows on Molecular Clouds}
\label{Impact of Outflows on Molecular Clouds}

In this section, we place our energy estimations (Section~\ref{Physical Properties of Outflows}) in the context of prior work and discuss the broader implications of the global impact of outflows. Most prior work derives the total outflow energy by assuming a moderate inclination angle and multiplying the 1D energy estimate by a factor of 3. For example, \citet{2014ApJ...783...29D} adopted an inclination angle of 57$^{\circ}.$3 which yields a energy correction factor of 3.4. Here we make the same assumption and report the 3D total, where we multiply our 1D estimate by a factor of 3.

To investigate the impact of outflows on turbulence, a typical approach is to compare the turbulence dissipation rate and the outflow energy injection rate \citep[e.g.,][]{2010ApJ...715.1170A,2015ApJS..219...20L,2020ApJ...896...11F}. However, the outflow energy injection timescale is highly uncertain. As discussed in Section~\ref{Correlation Between the Outflow Properties and the Number of YSOs}, we find the outflow mass is linearly correlated with the total number of YSOs instead of only the younger sources, which indicates that the driving timescale of outflows is even less clear. Instead, we compute the total outflow kinetic energy -- a quantity estimated as part of previous analyses in the determination of the outflow injection rate -- and compare our new values to those in the prior published studies. This allows us to conclude whether the impact of outflows is relatively similar to, more or less than that in prior analyses. 

\citet{2011ApJ...726...46N} conducted observations of \co\ (3-2) and \co\ (1-0) towards an active cluster-forming clump in Ophiuchus and identified six molecular outflows in both data. \citet{2011ApJ...726...46N} derived the outflow mass from \co\ (3-2) and \co\ (1-0) emission located in the high-velocity line wings. They derived a total outflow kinetic energy of $6\times10^{44}$ ergs. After comparing the total outflow energy injection rate with the dissipation rate of the supersonic turbulence, \citet{2011ApJ...726...46N} concluded that outflows inject significantly more energy than needed to offset the dissipation of turbulence. In our work, we find that the total outflow kinetic energy is $9\times10^{45}$ ergs, which is an order of magnitude larger than the value in \citet{2011ApJ...726...46N}. A larger area coverage in our work and using the optically thin tracer \13co\ in mass estimates might explain this difference. \co\ (3-2) and \co\ (1-0) are likely optically thick, which could cause their outflow mass, momentum and energy to be underestimated. More importantly, our method does not discard the outflow material ejected perpendicular to the line of sight that is located near the cloud central velocity. Since our energy calculation leads to a higher energy estimate, our result confirms the conclusion that outflow kinetic energy is sufficient to compensate for the dissipation of turbulence in Ophiuchus.

\citet{2015ApJS..219...20L} conducted a feedback census in Taurus and identified 55 outflows and 37 bubbles. They derived a total outflow energy of $3.9\times10^{45}$ ergs, which is 1\% of the Taurus cloud turbulent energy. However,  \citet{2015ApJS..219...20L} found that bubbles dominate the feedback energy. When also combining the kinetic energy from bubbles, they found the total kinetic energy from stellar feedback in Taurus is 30\% of the cloud turbulent energy. Since the energy injection rate from stellar feedback is comparable to the dissipation rate of the cloud turbulence, they concluded that stellar feedback, mainly in the form of bubbles, is sufficient to maintain turbulence in the current epoch. However, \citet{2020ApJ...890...64X} showed that the \citet{2015ApJS..219...20L} bubble energy estimate was substantially overestimated due to LOS contamination, which is caused by gas emission that is not associated with feedback being included in the outflow/bubble estimate. After correction, the kinetic energy from bubbles decreased by a factor of four. In our current work, we find the total outflow kinetic energy is $5.4\times10^{46}$ ergs, which is an order of magnitude larger than the value in \citet{2015ApJS..219...20L}. This difference is likely caused by a similar reason to that in Ophiuchus, i.e., our method includes the outflow material located around the cloud central velocity that is excluded in traditional approaches. Our calculation indicates that the kinetic energy from outflows is 14\% of the cloud turbulent energy. If we combine the contribution from bubbles \citep{2020ApJ...890...64X}, the total kinetic energy from stellar feedback is 22\% of the cloud turbulent energy. \citet{2015ApJS..219...20L} concluded that the outflow energy injection rate is marginally comparable to the dissipation rate of turbulence in Taurus, although outflow energy is only 1\% of the Taurus cloud turbulent energy. With our updated feedback energy, which is 22\% of the cloud turbulent energy, we conclude that the feedback energy injection rate can compensate for the turbulent dissipation rate \citep{2015ApJS..219...20L,2020ApJ...890...64X}.

\citet{2010ApJ...715.1170A} identified 60 outflow candidates in Perseus and derived a total outflow energy of $2\times10^{46}$ ergs, which is 12.5\% of the total cloud turbulent energy. However, it is worth noting that the method of calculating cloud turbulent energy used in \citet{2010ApJ...715.1170A} is different from our method. \citet{2010ApJ...715.1170A} adopted an average line width of 2 km/s to calculate the turbulent energy. While in this work, we calculate the turbulent energy channel by channel followed by Equation~\ref{energy-eq1}. The cloud turbulent energy derived by \citet{2010ApJ...715.1170A} is $1.6\times10^47$ ergs, while our approach indicates an estimate of $8\times10^47$ ergs, which is a factor of 5 larger than the estimate by \citet{2010ApJ...715.1170A}. To make a fair comparison, we adopt our turbulent energy estimate to discuss the impact of outflows in Perseus. Under these circumstances, the kinetic energy of the previously identified outflows are only 2.4\% of the total cloud turbulent energy. \citet{2020ApJ...905..172X} applied \CASItD\ to the same outflow catalog and predicted a total outflow energy of $7.8\times10^{45}$ ergs, which is one fourth of the value in \citet{2010ApJ...715.1170A}. This is caused by the lower mass-weighted LOS velocity as discussed in Section~\ref{Correlation Between the Outflow Properties and the Number of YSOs}. In our present work, we derive a total outflow energy of $1.2\times10^{47}$ ergs, which is 6 times larger than that in \citet{2010ApJ...715.1170A}. There are several possible reasons for this higher value. First, there are newly identified isolated outflows as discussed in \citet{2020ApJ...905..172X}. It should be noted that \citet{2020ApJ...905..172X} just followed up the 60 previous outflows and identified a few new outflows in the vicinity of these, whereas our current work make a prediction using the full CO map, which includes the entire cloud and a couple of clusters. This likely increase the outflow kinetic energy by a factor of a few. Second, \CASItD\ is able to identify all feedback structures in clustered regions, which are not fully identified in previous work. This also may enlarge the outflow energy by a factor of a few. Moreover, outflows identified by \citet{2010ApJ...715.1170A} are all distinct outflows with obvious coherent high-velocity structures that are likely powered by strong driving sources. However, in our work, \CASItD\ also identifies less distinct outflows whose velocity is small compared to previously identified ones. These weaker, less distinct outflows might be driven by more evolved or lower mass sources. This indicates that \CASItD\ provides a more inclusive outflow sample. 
\citet{2010ApJ...715.1170A} concluded that the 60 outflow candidates they identified play an important role in maintaining turbulence at the current epoch in Perseus, although the kinetic energy of outflows is 12.5\% of the total cloud turbulent energy in their study, or 2.4\% of the total cloud turbulent energy in our turbulence calculation. However, our updated outflow kinetic energy estimate is 15\% of the cloud turbulent energy. This indicates that outflows still play an important role in maintaining turbulence in Perseus.


\citet{2020ApJ...896...11F} identified 45 outflows near Herschel Orion Protostar Survey (HOPS) protostars \citep{2016ApJS..224....5F} in Orion but skipped the OMC-1 region. They derived a total outflow energy of $0.7-1.7\times10^{46}$ ergs. They found that the total outflow energy injection rate is comparable to the dissipation rate of turbulence. In our work, we derive a total outflow energy of $2.1\times10^{47}$ ergs, which is an order of magnitude larger than previous estimates. The main reason for the difference is that our survey covers a larger area and does not only focus on gas near the position of YSOs. In particular, we include the most active star forming region OMC-1. \CASItD\ predicts the kinetic energy from outflows in OMC-1 is $7.7\times10^{46}$ ergs, which is 37\% of the entire outflow energy in Orion. Thus, we confirm the prior conclusion that outflows are sufficient to maintain turbulence in Orion. 


In conclusion, our outflow kinetic energy estimates for all four clouds are larger than previous estimates. Consequently, we confirm the previous conclusion that outflows are an important agent to maintain turbulence at the current epoch. 

\subsection{Uncertainties of Outflow Estimates}
\label{Uncertainties of Outflow Estimates}

In Section~\ref{Physical Properties of Outflows}, we calculate the masses and dynamical properties of the outflows for the four clouds. In this section, we attempt to quantify the uncertainty of these outflow estimates. 

First, we assume a moderate excitation temperature of 25 K or the \co\ peak temperature, whichever is higher, to calculate the outflow mass. Most studies adopt values of excitation temperature in the range of 10-50 K \citep{2014ApJ...783...29D,2020ApJ...896...11F}. Since the mass scales linearly with the excitation temperature, the choice of excitation temperature introduces a factor of two uncertainty.

Meanwhile, \citet{2020ApJ...905..172X} found that model MF predicts the outflow mass within a scatter of 0.41. The most extreme offset case over-/under- estimates the mass by a factor of two. {We did not find any offset that would cause a systematic over or under-estimation. Meanwhile, we carry out one additional test to constrain the degree of possible contamination from cloud emission at rest-frame gas velocities. Appendix~\ref{CASItD Prediction on a New Synthetic Observation} shows an analysis of a more complex synthetic observation, which contains a number of interacting outflows. We conclude our \CASItD\ model is generally able to exclude contamination from the ambient gas near the rest frame velocity and may even underestimate the outflow contribution from low velocities in complex star-forming areas.
}

Furthermore, high-velocity, low-density gas emission is easily missed in high-velocity channels due to the low signal-to-noise. \citet{2020ApJ...905..172X} used the simulations to estimate that 10\% of the outflow gas is lost. The correction factors for LOS momentum and LOS energy are 1.3 and 1.8 \citep{2020ApJ...905..172X}, respectively. 

In addition, the inclination angle of the outflow plays an important role in converting LOS momentum/energy to 3D momentum/energy. In this work, we adopt correction factors of $\sqrt{3}$ and 3 to convert LOS momentum and LOS energy to 3D estimates. These correspond to an inclination angle of 55$^{\circ}$. If the inclination angle is between 20$^{\circ}$ and 70$^{\circ}$, this leads to a factor of 2 uncertainty for 3D momentum and a factor of 4 uncertainty for 3D energy.

Finally, chemical conditions, i.e., \co\ and \13co\ abundances, have an influence on the mass estimate. In our mass calculation, we adopt constant abundance ratios, [\co]/[\h2]=$10^{-4}$ and [\co]/[\13co]=62. However, these abundance ratios vary across diffident clouds. We cannot evaluate this uncertainty without sophisticated chemical modeling for each cloud. 

Consequently, by combining these uncertainties and using error propagation, we find a factor of 3 uncertainty for mass, a factor of 3.4 uncertainty for 3D momentum, and a factor of 5 uncertainty for 3D energy.

\section{Conclusions}
\label{Conclusions}

We apply the deep learning method \CASItD\ to four nearby molecular clouds, Ophiuchus, Taurus, Perseus and Orion, to systemically identify protostellar outflows and study the impact of outflows on their host clouds. Our main findings are the following:

\begin{enumerate}

\item The total outflow masses are 267 \msun, 795 \msun, 1305 \msun\ and 6332 \msun\ for Ophiuchus, Taurus, Perseus and Orion, respectively. On average, the mass associated with feedback is around 10\% of the host cloud mass for all four clouds.

\item The total 3D outflow energies are $9\times10^{45}$ ergs, $6\times10^{46}$ ergs, $1.2\times10^{47}$ ergs and $6\times10^{47}$ ergs for Ophiuchus, Taurus, Perseus and Orion, respectively. 

\item The outflow mass is linearly proportional to the total number of YSOs for all four clouds. On average, each YSO is associated with one solar mass of outflow material. 

\item The outflow momentum is linearly proportional to the total number of YSOs for all four clouds. On average, each outflow has a mass-weighted LOS velocity of 1 km/s. The relatively low value is because more mass is located near the cloud central velocity. 

\item We compute the spatial power spectrum of the outflow prediction map. We find all four clouds exhibit a break point, which ranges from 0.27 pc (Orion) to 0.65 (Perseus) pc. The break point likely indicates the typical outflow mass and energy injection scale. 


\item Models trained to identify outflows are likely to also identify bubble structures, which may be produced by main sequence massive stars. 

\item We compare the energy associated with outflows to the rate of turbulent dissipation and conclude that feedback is sufficient to maintain turbulent dissipation at the current epoch for all four clouds. 

\end{enumerate}

D.X., S.S.R.O., and R.A.G. acknowledge support by NSF grant AST-1812747. D.X. acknowledges support from David Alan Benfield Memorial Scholarship in Astronomy and from the Virginia Initiative on Cosmic Origins (VICO). S.S.R.O.  acknowledges support from NSF Career grant AST-1748571. R.A.G. acknowledges support from NASA ADAP grant NNX17AF24G. HGA acknowledges support from NSF grant AST-1714710. The Texas Advanced Computing Center (TACC) at the University of Texas at Austin provided HPC resources that have contributed to the research results reported within this paper. 

\appendix

\section{Discussion on the Effect of Box Sizes on the YSO-Mass Relation}
\label{Discussion on the Effect of Box Sizes on the YSO-Mass Relation}

In this section, we examine the effect of our choice of ``window'' sizes when studying the correlation between the total mass associated with outflows and the number of YSOs. We define windows with physical sizes of 0.5~pc$\times$0.5~pc, 1~pc$\times$1~pc, 2~pc$\times$2~pc, 3~pc$\times$3~pc and 5~pc$\times$5~pc to scan through the entire cloud region. We set a step size such that each box overlaps with its neighbors by at least 80\%. Then we calculate the mass inside the window and count the number of YSOs enclosed. We consider the outflow relationship between the number of both young YSOs and total YSOs.

\begin{figure*}[]
\centering
\includegraphics[width=0.98\linewidth]{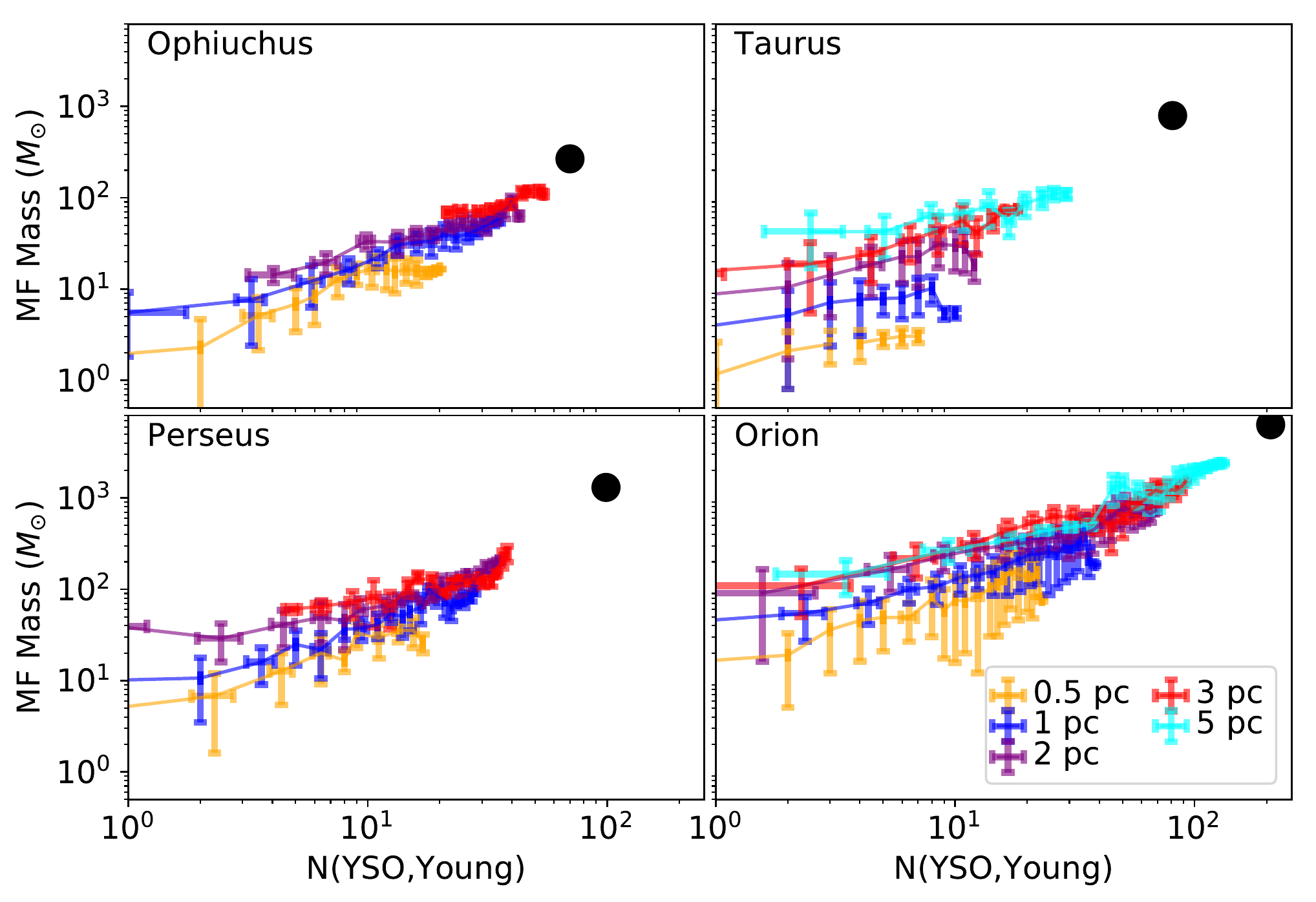}
\caption{Correlation between the mass associated with feedback and the number of young YSOs. The different colors indicate different window sizes for the estimation. The filled circles indicate the total outflow mass and the total number of young YSOs in the four clouds. }
\label{fig.box-size-yso-mass-mf-young}
\end{figure*}

Figure~\ref{fig.box-size-yso-mass-mf-young} shows the correlation between the model MF outflow prediction and the number of young YSOs for the five window sizes applied to the four clouds. For Ophiuchus, Perseus and Orion, when the window size is above certain threshold, the correlations between outflow mass and the number of YSOs become similar, which indicates the choice of window size does not significantly change the results. Taurus is more sensitive to the window size than the other regions. This is likely caused by the large separations between YSOs. However, it still shows a clear trend of smaller separation between different correlations when the window size gets larger.


The trends are similar when considering total YSOs. Consequently, we conclude that the study of the correlation between the outflow properties and the number of YSOs is robust.

\section{Spectra of Outflows and Bubbles}
\label{Spectra of Outflows and Bubbles}

In this section, we employ an unsupervised learning algorithm t-SNE (t-distributed Stochastic Neighbor Embedding) to compare the spectra identified as outflow feedback to those identified as belonging to bubbles. t-SNE is a tool to visualize high-dimensional data by converting the similarities between data points into a low-dimensional manifold. It has been used for classification and outlier detection in a variety of astronomy data \citep{2017MNRAS.472.2517J,2018MNRAS.476.2117R,2020arXiv201011202L,fluke2020surveying}. For example, \citet{2018MNRAS.476.2117R} successfully applied t-SNE to cluster stars based on their spectra and identify outliers based on the t-SNE map.

We take the L1688 star cluster region in Ophiuchus as an example. Figures~\ref{fig.pv-map-oph-goodcase} and \ref{fig.pv-map-oph-bubble} show the prediction of this star cluster region by models trained to identify only outflows and only bubbles, respectively. In some cases the models clearly identify only one or the other type of feedback, however, there is also significant overlap. In order to classify the spectra, we calculate outflow mass and bubble mass of each spectra. If the outflow mass or bubble mass for a given line of sight is above 5$\sigma$ compared with the background noise, we consider that spectrum to contain feedback. If the outflow mass is larger than the bubble mass, we label it an outflow spectrum. If the bubble mass is larger than the outflow mass, we label it a bubble spectrum. If there is emission but neither outflow mass nor bubble mass exceeds the 5$\sigma$ threshold, or the spectrum has less than three channels that contain feedback, we label this spectrum as non-feedback. We neglect any spectra that have no emission above 3$\sigma$ sigma.

\begin{figure*}[hbt!]
\centering
\includegraphics[width=0.98\linewidth]{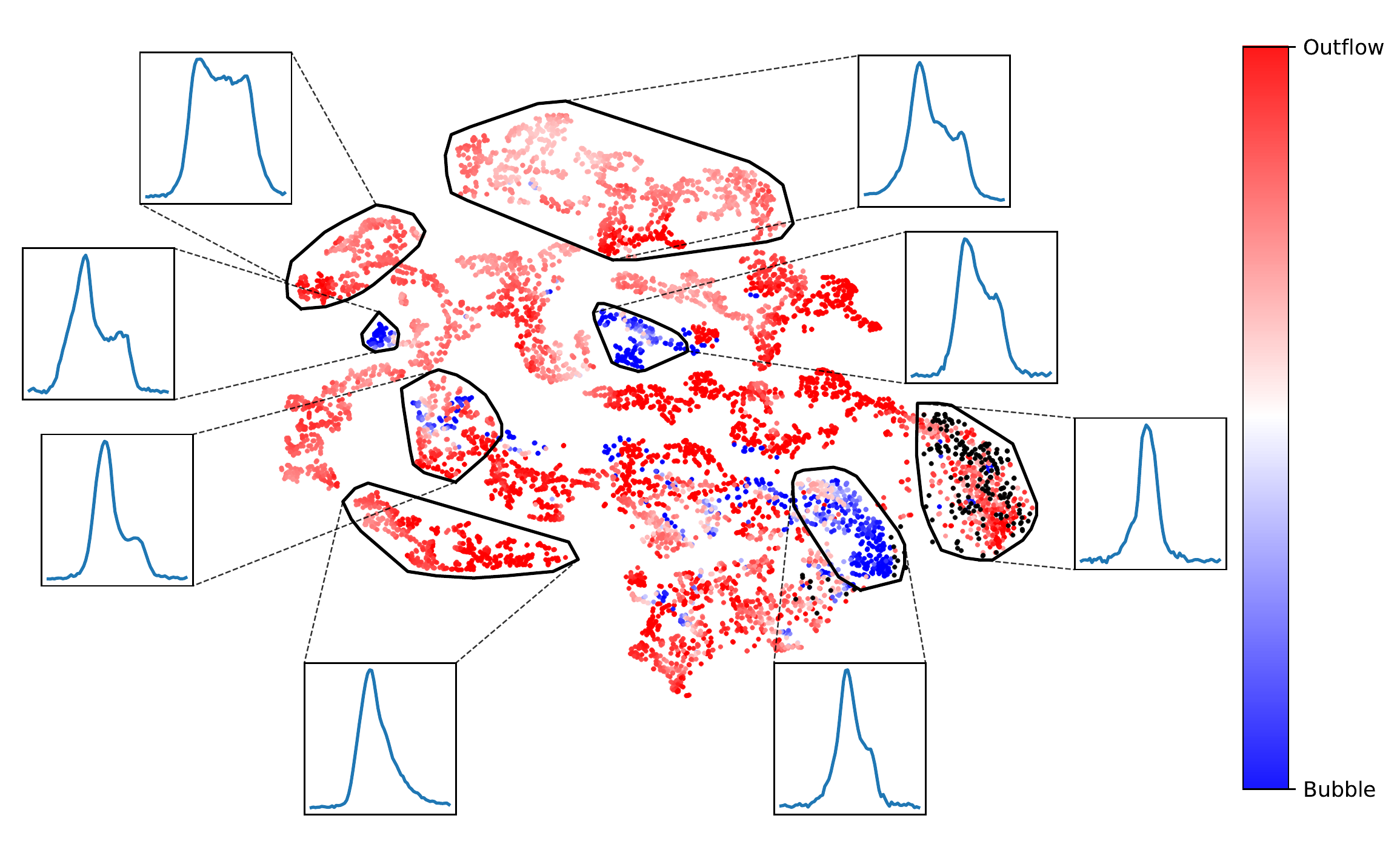}
\caption{t-SNE map of the \co\ spectra in the L1688 star cluster region in Ophiuchus. Red dots indicate outflow spectra. Blue dots indicate bubble spectra. Black dots indicate non-feedback spectra. The sub-plots show the average spectra of each subregion.}
\label{fig.tsne-oph-b2-spec}
\end{figure*}

{We use the spectra of the L1688 region at each line of sight as input to t-SNE}. 
{t-SNE projects the higher dimensional data into a lower dimensional (2D) space and organizes it based on similarities (e.g., the line shape and intensity of a spectrum) in the high-dimension space. }
Figure~\ref{fig.tsne-oph-b2-spec} shows the t-SNE clusters. {We color the data points based on the \CASItD\ predictions,} where spectra predicted to be primarily associated with outflows are in red, those primarily associated with bubbles are in blue and non-feedback in black. It is clear that the spectral shape for bubbles and outflows are different. The spectra of bubbles show two Gaussian components with two peaks of different intensities. While the spectra of outflows have various shapes, e.g., broader line wings, two Gaussian components but with almost equal peaks. It's worth noting that broad line wings are often used to visually identify outflows\citep{2015ApJS..219...20L,2019PASJ...71S...8T}. t-SNE separates some bubble spectra from the outflow spectra. Meanwhile, we notice that there are some blended/transiting regions, where outflow spectra and bubble spectra are clustered together. In these cases, we find that bubble and outflow identifications have similar spectral shapes. This demonstrates that we cannot distinguish outflows and bubbles based only on single spectra. There is also a mixed region where non-feedback spectra are blended with outflow spectra. The spectral shape of this region indeed show less outflow features, where there is a faint broad line-wing on the blue shifted side.

This analysis explores the similarity between outflow and bubble feedback {using an unsupervised machine learning approach}. In some cases, the \CASItD\ predictions are clearly distinct. In other cases, where both models identify feedback emission, it is clear that the spectra are truly similar. This may be because both types of feedback can produce such velocity features or because such regions genuinely contain both types of feedback. In the current analysis it is not possible to distinguish between these two possibilities. However, this t-SNE clustering gives confidence that feedback is shaping these spectra and that \CASItD\ is correctly identifying feedback signatures.

\section{YSOs in Orion}
\label{YSOs in Orion}

\begin{figure*}[hbt!]
\centering
\includegraphics[width=0.88\linewidth]{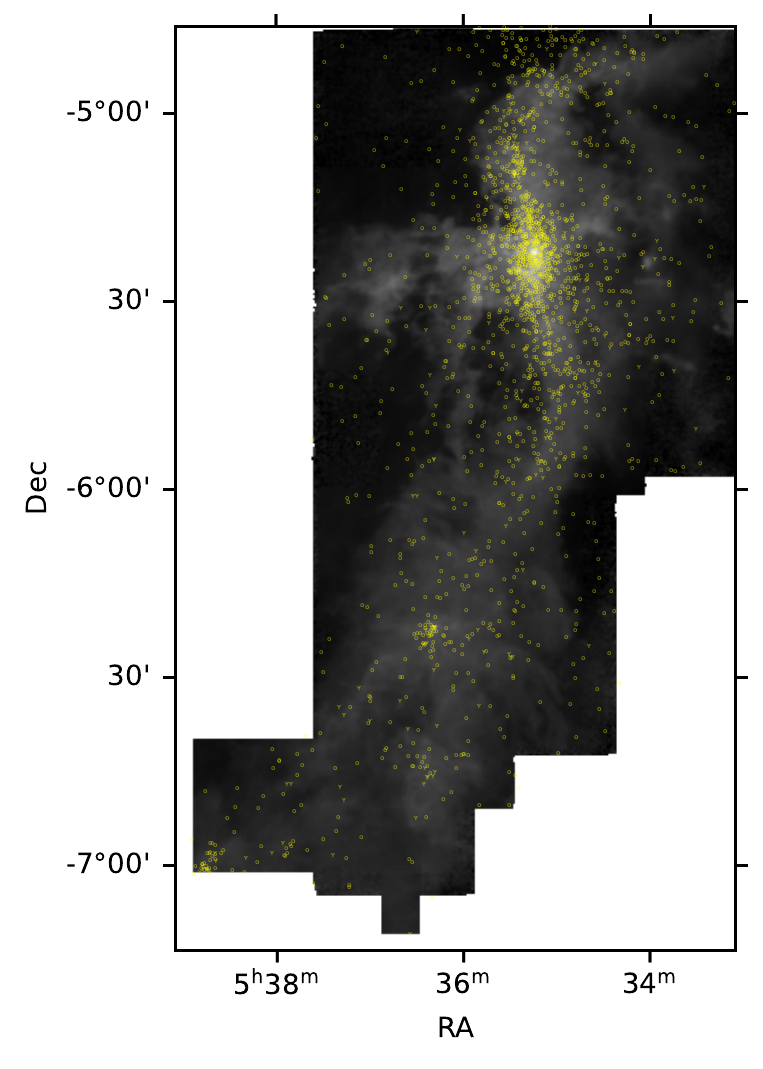}
\caption{Intensity of \co\ integrated over all velocity channels for Orion. Letters ``Y'' and ``O'' mark YSO positions, as described in Section~\ref{YSO Catalogue}.}
\label{fig.yso-orion-lm-all}
\end{figure*}

Orion has an extensive population of older YSOs, so in Figure~\ref{fig.pred-orion-lm-ME1} we plot only the young YSOs for readability. Here, Figure~\ref{fig.yso-orion-lm-all} shows the location of both young and old YSOs in Orion. While we expect younger YSOs to contribute more feedback per source, the large number of older YSOs indicates that they likely contribute feedback throughout the cloud, and this is consistent with the prediction in Section~\ref{Orion}.


\section{Outflows in $^{13}$CO }
\label{Outflow Morphology}

\begin{figure*}[hbt!]
\centering
\includegraphics[width=0.98\linewidth]{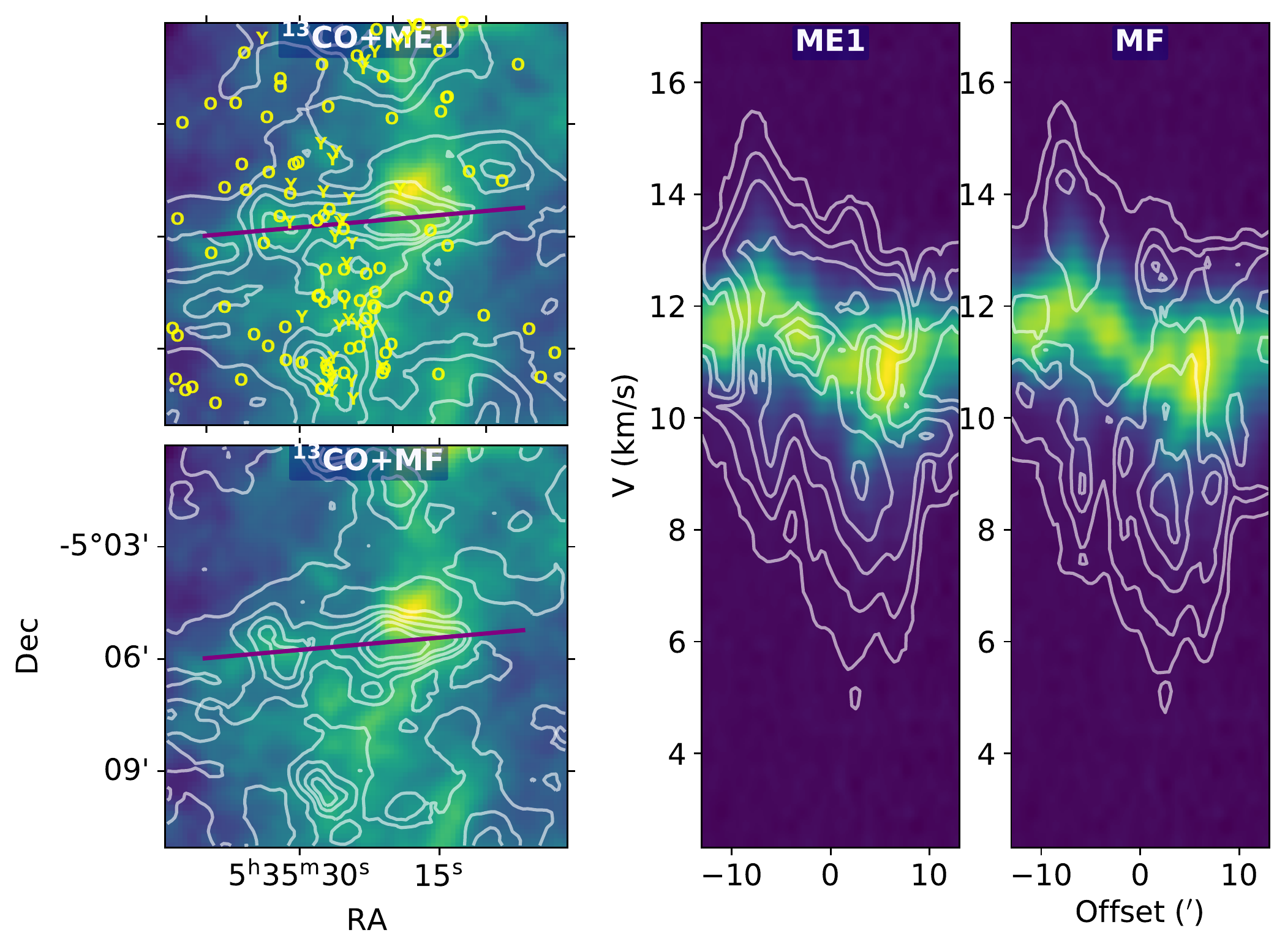}
\caption{Same as Figure~\ref{fig.pv-map-orion-goodcase} but {where the background colorscale indicates the \13co\ integrated intensity. White contours show the prediction by models ME1 and MF on the \co\ data. 
}}
\label{fig.pv-map-orion-13cocase}
\end{figure*}

In some regions \co\ is optically thick and we combine both \co\ and \13co\ to derive the kinematic estimates rather than using \co\ only \citep[e.g.,][]{2010ApJ...715.1170A}. In most cases, the velocity channels where there is \13co\ emission dominate the mass in each spectrum. Here we compare the identification from the \co\ map to the \13co\ emission. Figure~\ref{fig.pv-map-orion-13cocase} shows the \13co\ emission of the outflow in Figure~\ref{fig.pv-map-orion-goodcase}. We can clearly see the velocity range of \13co\ is substantially smaller than that of \co. However, the mass in the \13co\ emitting region dominates the total mass of the cloud. This also explains why the mass weighted LOS velocity is smaller than the velocity calculated based on \co\ only.

{ 

\section{Correlation Between the Gas Mass and the Number of YSOs}
\label{Correlation Between the Gas Mass and the Number of YSOs}

In this appendix, we examine the relationship between the number of YSOs, the total gas mass, and the dense gas mass. Prior work has shown that the number of YSOs is well-correlated with the gas column density, which is $\Sigma_{*}\propto \Sigma_{\rm gas} ^{2}$, or $M_{\rm gas} \propto N_{*} ^{0.5}$ \citep[e.g.,][]{2011ApJ...739...84G,2020ApJ...896...60P,2021ApJ...912L..19P}, where the gas mass was estimated from dust emission. \citet{2020ApJ...896...60P,2021ApJ...912L..19P} found that the surface density of protostars (i.e., younger YSOs) are tightly correlated with gas column following the relation above. \citet{2011ApJ...739...84G} found a similar but less tight relation when considering all YSOs. Consequently, we expect YSOs and dense gas to be correlated. 
Figure~\ref{fig.correlation-yso-mass} shows a nearly linear relation between the total mass and the number of all YSOs. We find the slope is 0.93, which is slightly smaller than the slope of the outflow mass-YSO relation, 1.03. 

To check whether the outflow mass estimate is correlated with the dense gas due to contamination, we examine the relation excluding the rest-frame velocity gas, which is located within the turbulent velocity range ($\pm 1$ km/s). Figure~\ref{fig.correlation-yso-vcut-mass} shows the correlation between the total high velocity gas ($|v-v_{cen}|>1$ km/s) mass and the number of all YSOs, and the correlation between the high velocity outflow mass ($|v-v_{cen}|>1$ km/s) and the number of all YSOs. The slopes are similar to that of the total gas/outflow gas relation including the rest-frame velocity gas. Therefore, we conclude the correlation between gas mass and YSOs exists because outflows are linearly proportional to the number of YSOs launching them. YSOs form from dense gas and thus by nature are more likely to be located in denser regions.  By focusing only on the high-velocity material, which traces the outflows most directly, we show that the trend arises directly from the expected correlation between YSOs and outflows rather than indirectly from contamination by the dense gas. This result provides further evidence that our method performs well in dense regions where contamination from the cloud would otherwise be a serious problem.

\begin{figure}[hbt!]
\centering
\includegraphics[width=0.48\linewidth]{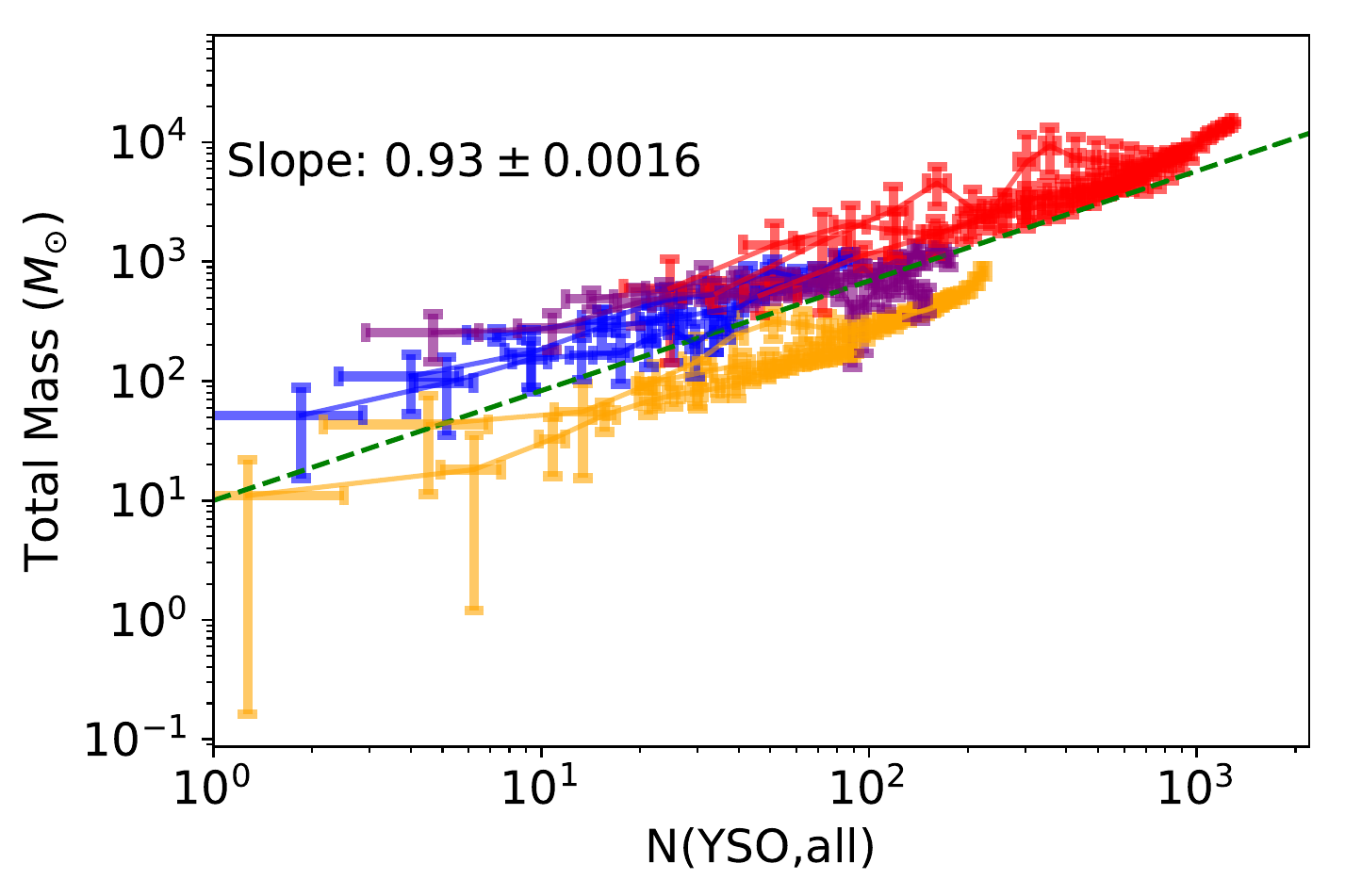}
\includegraphics[width=0.48\linewidth]{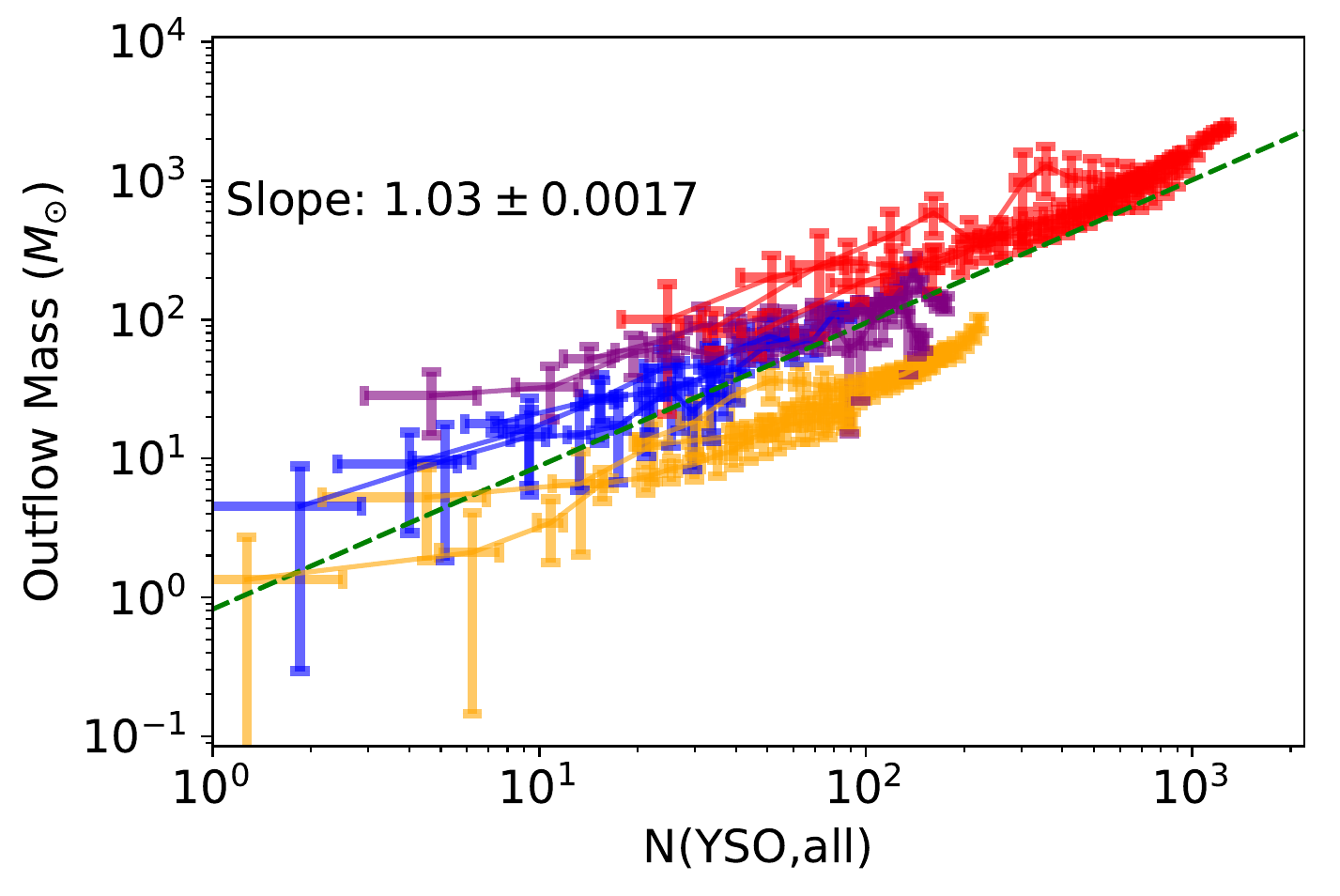}
\caption{Left: correlation between the total mass and the number of all YSOs. Right: correlation between the outflow mass and the number of all YSOs.  }
\label{fig.correlation-yso-mass}
\end{figure}

\begin{figure}[hbt!]
\centering
\includegraphics[width=0.48\linewidth]{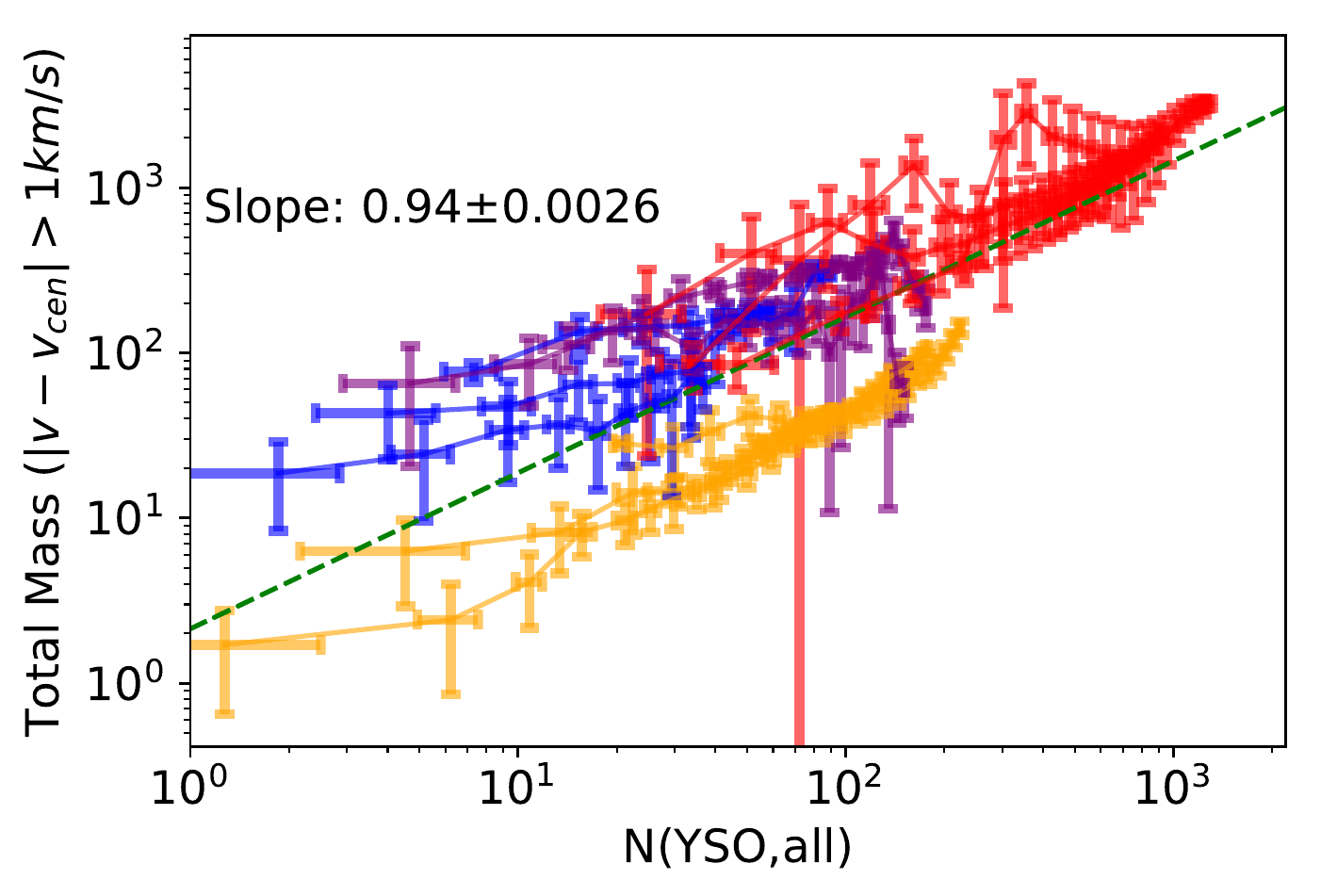}
\includegraphics[width=0.48\linewidth]{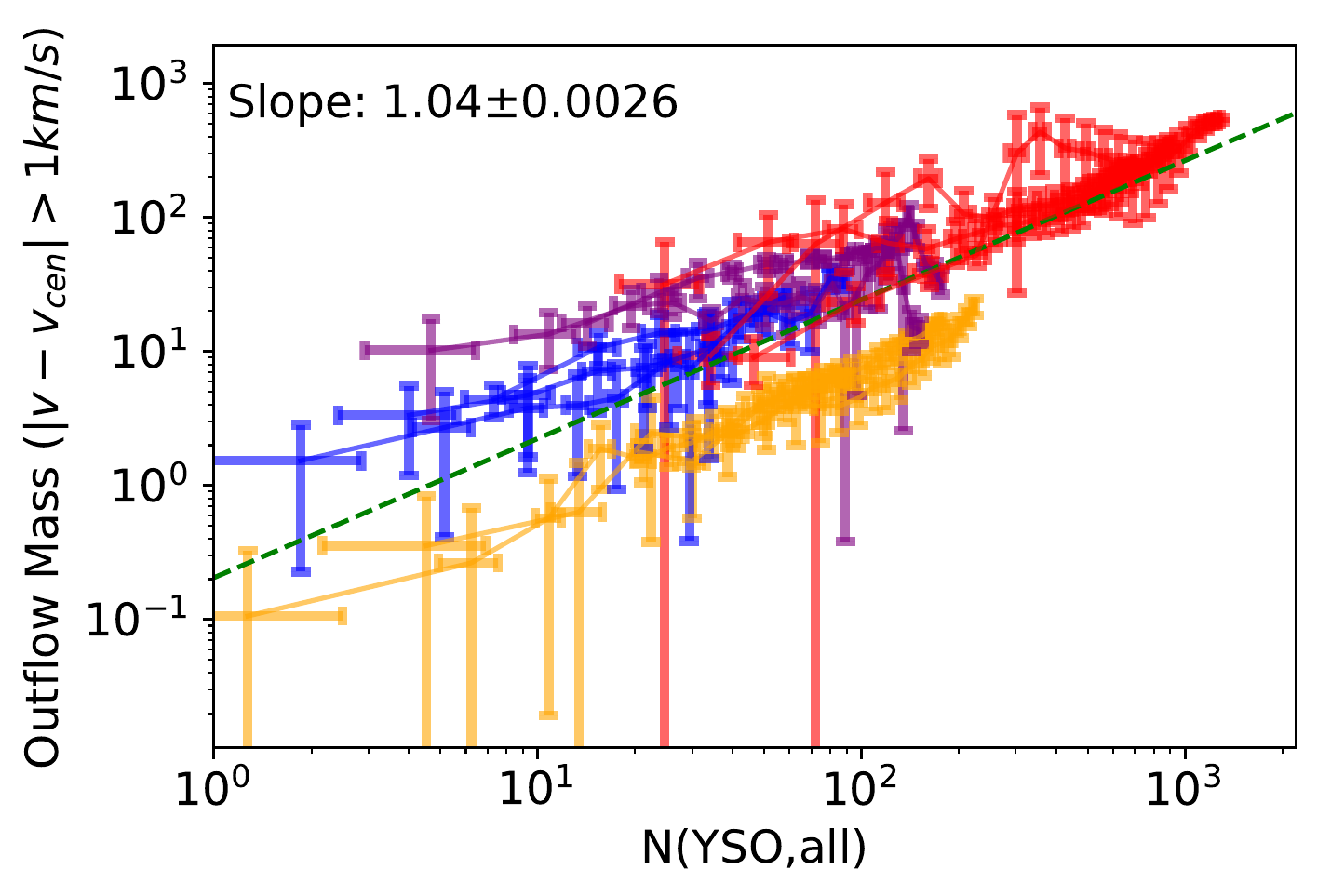}
\caption{Left: correlation between the total high velocity gas ($|v-v_{cen}|>1$ km/s) mass and the number of all YSOs. Right: correlation between the high velocity outflow mass ($|v-v_{cen}|>1$ km/s) and the number of all YSOs.  }
\label{fig.correlation-yso-vcut-mass}
\end{figure}

\section{\CASItD\, Prediction on a Complex Synthetic Observation}
\label{CASItD Prediction on a New Synthetic Observation}

In this appendix, we examine a prediction by the two \CASItD\, models on a synthetic observation of a simulation snapshot with a number of interacting outflows. This synthetic observation and the simulation it derives from is not included in our previous training or testing, so it constitutes a more challenging test of the model performance. This simulation is run using a different initial turbulent seed and with twice the inital gas density as the other simulations in the training set. The mean magnetic field strength is 0.8 $\mu$G. The evolutionary time of the snapshot is 0.7 $t_{\rm ff}$. There are 13 stars launching outflows in the simulation box, whereas our training set is constructed from snapshots with 1 to 5 stars. 
Consequently, this output provides an independent check on the \CASItD model performance.

Figure~\ref{fig.synth-outflow-channel} shows the integrated CO emission, tracer fractions and predictions. 
We find the \CASItD\ model does not actually over-predict the amount of emission in the cloud velocity channels but instead slightly under-predicts the location and mass of the outflows. More quantitatively, model MF identifies 76\% of the outflow emission at high-velocity channels (above 2\kms). However, model MF only identifies 40\% outflow emissions near the rest-frame velocity (within 2 \kms). This implies \CASItD\ might underestimate the outflow mass by a factor of 2. This is consistent with the model uncertainty discussed in Section~\ref{Uncertainties of Outflow Estimates}. This analysis gives further confidence that \CASItD\ correctly excludes contamination from the ambient gas near the rest frame velocity, even in relatively clustered and complex regions.

\begin{figure}[hbt!]
\centering
\includegraphics[width=0.98\linewidth]{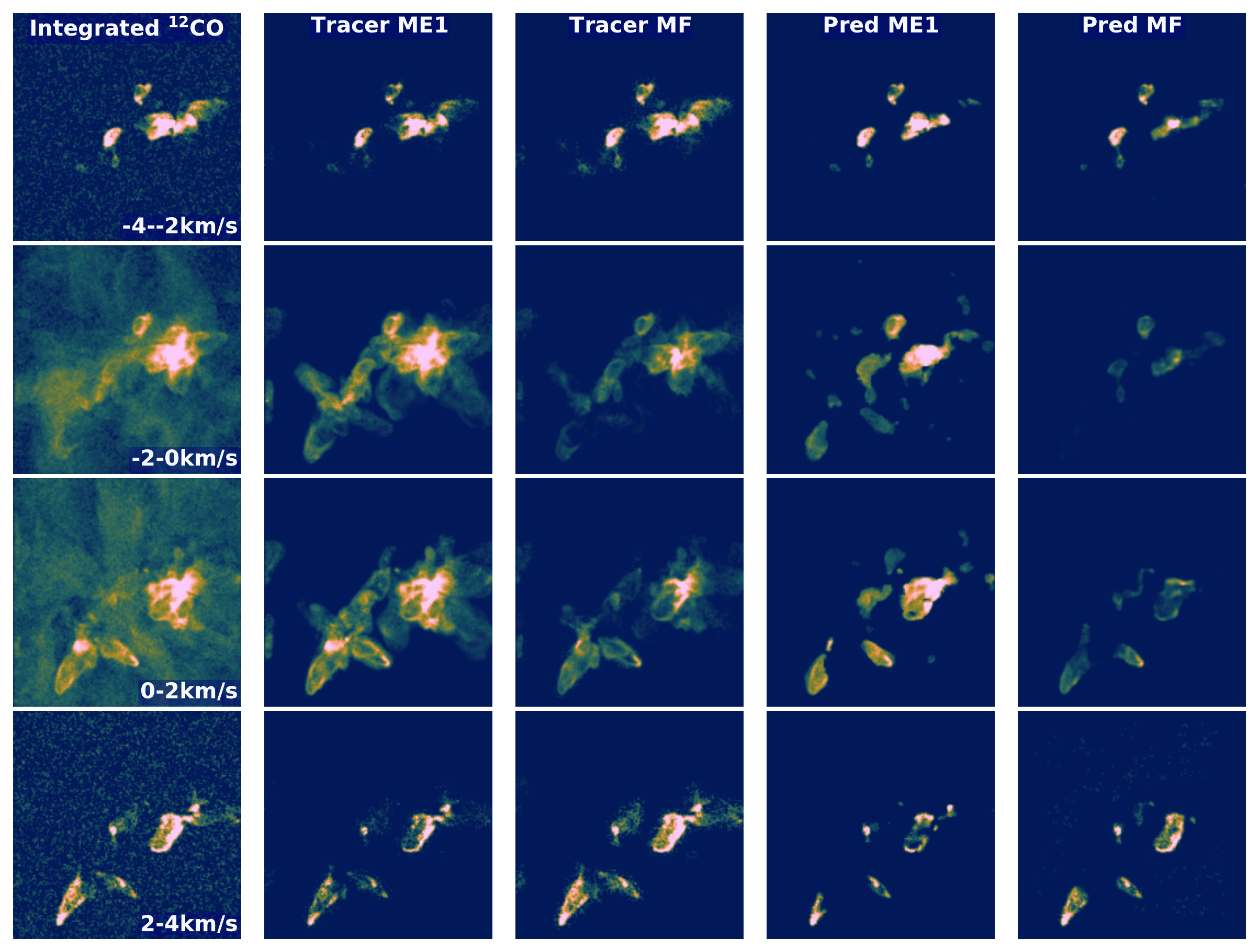}
\caption{Synthetic \co\ observations and \CASItD\ prediction in different velocity channels. }
\label{fig.synth-outflow-channel}
\end{figure}

}

\bibliographystyle{aasjournal}
\bibliography{references}

\end{CJK*}

\end{document}